\documentclass[preprint,10pt,authoryear]{elsarticle}

\usepackage{graphicx}
\usepackage{amsmath,amssymb,bm}
\usepackage{subcaption}           
\usepackage{scalerel}
\usepackage{hyperref}
\usepackage{epstopdf}
\usepackage{epsfig}
\usepackage[font=small,labelfont=bf,format=plain,justification=justified]{caption}

\usepackage{siunitx}
\usepackage{multirow}

\hypersetup{
    colorlinks = true,
    linkcolor  = blue,
    urlcolor   = blue,
    citecolor  = blue,
}

\newcommand{\RomanNumeralCaps}[1]{} 

\usepackage{xcolor}

\journal{International Journal of Multiphase Flow}

\begin{document}
\begin{frontmatter}

\title{ Particle Thermal Inertia Delays the Onset of Convection in Particulate Rayleigh–Bénard System}

\author[aff1]{Saad Raza\corref{cor1}}
\ead{saad.raza@univ-lille.fr}

\author[aff1]{Apolline Lemoine}
\author[aff1]{Yan Zhang}
\author[aff1]{Enrico Calzavarini}
\author[aff2]{Romulo~B. Freitas}
\author[aff3]{Leonardo~S.~de~B. Alves}
\author[aff1]{Silvia~C. Hirata}

\cortext[cor1]{Corresponding author.}

\address[aff1]{Univ.~Lille, Unit\'e de M\'ecanique de Lille -- J.~Boussinesq (UML) ULR 7512, F-59000 Lille, France}
\address[aff2]{Federal Center for Technological Education Celso Suckow da Fonseca, Nova Igua\c{c}u-RJ 26041-271, Brazil}
\address[aff3]{PGMEC, Univ. Federal Fluminense, Rua Passo da P\'atria 156, Niter\'oi, Rio de Janeiro 24210-240, Brazil}

\begin{abstract}
We investigate the linear stability of a thermally stratified fluid layer confined between horizontal walls and subject to continuous injection of dilute thermal particles at one boundary and extraction at the opposite, forming a particulate Rayleigh-B\'enard (pRB) system. The analysis focuses on the influence of thermal coupling between the dispersed and carrier phases, quantified by the specific heat capacity ratio $\epsilon$. Increasing $\epsilon$ systematically enhances stability, with this effect persisting across a wide range of conditions, including heavy and light particles, variations in volumetric flux, injection velocity and direction, and injection temperature. The stabilizing influence saturates when the volumetric heat capacity of the particles approaches that of the fluid, $\epsilon = \mathcal{O}(1)$. The physical mechanism is attributed to a modification of the base-state temperature profile caused by interphase heat exchange, which reduces thermal gradients near the injection wall and weakens buoyancy-driven motion.
\end{abstract}

\begin{keyword}
Natural convection \sep particle-laden flows \sep two-way coupling
\end{keyword}

\end{frontmatter}


\section{Introduction}

Particle-laden thermal convection is a phenomenon of broad scientific and technological significance, emerging in systems where heat transfer and buoyancy interact with dispersed solid, liquid, or gaseous phases. Such flows are encountered across a wide spectrum of engineering, environmental, and geophysical processes, from solar-thermal receivers and packed-bed heat exchangers to magma chambers, sediment plumes, and atmospheric convection \citep{ho2014review, ho2016review, de2008heterogeneities, weber2020magma, bergantz2015open, mingotti2022dynamics, ernst1996sedimentation, wright2001effects, Grabowski2013,patovcka2022residence}. In these systems, the presence of suspended particles or bubbles modifies both momentum and heat transport, often giving rise to new convective regimes, altered heat-flux scaling, and pattern transitions not observed in single-phase flows. In solar-energy applications, radiatively heated particle suspensions are used as efficient volumetric absorbers in high-temperature particle solar receivers and thermal-energy storage systems, where particle-fluid coupling governs both absorption efficiency and convective stability \citep{ho2016review, Yang_Wan_Zhou_Dong_2022}. Similarly, fluidized-bed reactors and granular heat exchangers rely on particle-laden convective transport for uniform temperature distribution and enhanced mixing \citep{Gidaspow1994, grace2012circulating}. In materials processing—including metal casting, additive manufacturing, and solidification control—the onset and structure of convection within the melt strongly influence microstructural patterning and defect formation \citep{glicksman1988circulating, zheng2023numerical}.

The stability of a quiescent fluid layer containing a dispersed phase of particles, drops, or bubbles is a classical problem in multiphase fluid dynamics, known for its richness and complexity \citep{prosperetti1987linear, Guazzelli2011,LegendreZenit2025}. Even under idealized conditions, its mathematical formulation involves numerous physical parameters describing the material properties of both the carrier and dispersed phases, as well as their boundary and injection conditions. The problem becomes more intricate when the coupling between phases is not only mechanical but also thermal, involving interphase heat transfer or, in more general cases, latent heat exchange due to melting, condensation, or evaporation. When the suspended particles are small, numerous, and sufficiently dilute, their collective effect on the carrier fluid can be modeled using an Eulerian two-fluid framework. Within this continuum approach, the dispersed phase interacts with the carrier through drag, buoyancy, and thermal coupling, allowing one to determine the parametric conditions under which particle motion either stabilizes or destabilizes the fluid layer, changing the onset of large-scale convection \citep{Saffman_1962}.

In recent decades a number of linear and weakly nonlinear stability analyses have revisited such systems, clarifying mechanisms by which the dispersed phase influences onset thresholds and pattern formation. For instance, \citet{NakamuraPhysRevE.102.053102, Nakamura_Yoshikawa_Tasaka_Murai_2021} studied bubble-induced convection in horizontal liquid layers via linear and bifurcation analyses, demonstrating how bubble buoyancy and drag modify the critical Rayleigh number. Similar efforts to include interphase coupling and thermal effects in multiphase stability analysis were made by \citet{ayukai2023derivation}, who developed and analyzed a two-fluid model for bubbly flows incorporating bubble oscillations, viscous damping, and heat transfer. \citet{Srinivas_Tomar_2025} extended the investigation of particle-laden Rayleigh–Bénard convection through a weakly nonlinear Landau‐type analysis, showing how settling particles and nonlinear mode interactions break symmetry and give rise to harmonic generation and preferential concentration. Recently, \citet{KangLee2025} performed a linear stability analysis of particle-laden Rayleigh–Bénard convection and found that the critical Rayleigh number increases with the product of the particle mass loading and the non-dimensional terminal velocity, indicating that settling small, heavy particles tend to stabilize the flow. More recently, the stability of the particulate Rayleigh–Bénard (pRB) system has been examined for various particle densities, encompassing both heavy and light dispersed phases \citep{razaPF2024}. In this configuration, a viscous fluid layer is confined between two horizontal plates held at different temperatures, the lower plate being hotter, which establishes an unstable density stratification. Particles are continuously introduced into the layer at their terminal velocity and at a prescribed temperature, entering through the wall they move away from—namely, the top wall for heavy particles that fall downward and the bottom wall for light particles that rise upward—and exiting at the other boundary. Linear stability analysis has shown that the presence of particles increases the critical Rayleigh number required for the onset of convection, indicating that the particulate system is more stable than the single-phase case. This stabilizing behavior persists regardless of the relative density of the two phases and the strength of the mechanical or thermal couplings. Similar results were also obtained for heavy particles settling from the top boundary, where convection was likewise suppressed \citep{prakhar2021linear}. In these studies, the injection velocity was fixed at the terminal value, enforcing a uniform particle concentration throughout the layer and preventing accumulation or depletion. A subsequent investigation relaxed this constraint by allowing the injection velocity to differ from the terminal one, exploring both sub-terminal and super-terminal regimes \citep{Raza_Freitas_Alves_Calzavarini_Hirata_2025}. In that work, the thermal coupling between the phases was retained but assumed weak, enabling focus on the mechanical effects of particle injection and sedimentation. It was shown that the injection velocity acts as a control parameter that can either stabilize or destabilize the system, depending on the particle density ratio and direction of motion. These results highlighted the key role of boundary injection conditions in determining convective stability in particulate Rayleigh–Bénard systems and motivated the present investigation into how energetic couplings, specifically the finite thermal capacity of the dispersed phase, affect the onset of convection.

Previous stability analyses have mainly focused on the mechanical aspects of particle–fluid coupling, whereas the influence of thermal processes has received comparatively little attention. The common assumption that particles instantaneously equilibrate thermally with the surrounding fluid is convenient but often unrealistic. Finite particle size, heat capacity, and thermal conductivity introduce a nonzero thermal relaxation time during which the particle temperature adjusts to that of the local fluid. This delay defines the particle thermal inertia, which controls the rate of interphase heat exchange. Thermal inertia introduces a phase lag between fluid and particle temperature fluctuations, altering the energy available to sustain convective motion. Depending on the ratio of thermal relaxation to the characteristic flow timescale, this effect can either enhance or suppress instability. In most previous pRB studies, the thermal coupling strength was expressed through the volumetric heat capacity ratio, which combines the effects of particle density and specific heat. In the present work, these contributions are decoupled to isolate the role of particle thermal inertia. A separate specific heat capacity ratio between the two phases, denoted $\epsilon$, is introduced to describe the intrinsic thermal response of the particles independently of their density ratio. This distinction separates the energetic effects of heat storage and exchange from the mechanical effects of buoyancy and momentum coupling, and it enables consistent modeling of cases such as gas bubbles in liquids, where the thermal coupling vanishes regardless of the particle heat capacity.

The objective of this study is to quantify the influence of finite particle thermal inertia on the stability of the particulate Rayleigh–Bénard system by varying the specific heat capacity ratio $\epsilon$, and thereby the thermal relaxation time. The analysis extends the linear stability framework developed in earlier work \citep{razaPF2024,Raza_Freitas_Alves_Calzavarini_Hirata_2025} to include the coupled evolution of fluid and particle temperatures with finite interphase heat transfer. Results show that particle thermal inertia consistently delays the onset of convection, increasing the critical Rayleigh number relative to both the particle-free and thermally equilibrated (purely mechanically coupled) pRB systems. The stabilizing mechanism originates from a modification of the fluid temperature base state due to heat exchange with suspended particles. When the coupling is strong, i.e. when $\epsilon \ge 1$, the temperature profile flattens near the particle injection region, reducing the thermal gradient that drives convection. The resulting convective rolls are confined toward the opposite wall and require stronger buoyant forcing to develop. Although thermal and mechanical couplings act simultaneously, their combined effect can lead to destabilization when light particles dominate the mechanical forcing. Overall, the analysis clarifies how finite particle thermal inertia modifies the energetic balance governing convective instability in particulate systems.

The paper is organized as follows. Section \ref{sec:model} presents the governing equations and the dimensionless parameters that define the particulate {Rayleigh-B\'enard} model. Section \ref{sec:stability} describes the base flow, the linear stability formulation and the corresponding eigenvalue problem. Results are discussed in Section \ref{sec:results}, emphasizing the influence of particle thermal inertia on the onset of convection. Concluding remarks are given in Section \ref{sec:conclusion}.

\section{\label{sec:model}The particulate Rayleigh-B\'enard model system}

We adopt an Eulerian framework to describe the dynamics of a dilute suspension of macroscopic particles in the pRB configuration~\citep{Raza_Freitas_Alves_Calzavarini_Hirata_2025}. The particle volume concentration is assumed small, allowing the carrier fluid to be treated as incompressible and governed by the Navier–Stokes equations under the Boussinesq approximation for the velocity $\textbf{u}(\bm{x},t)$ and temperature $T(\bm{x},t)$ fields. Owing to conservation of total momentum and thermal energy, the particulate phase exerts both mechanical and thermal feedback on the fluid. Each particle is characterized by its material properties: density $\rho_p$, diameter $\textrm{d}p$, and specific heat capacity at constant pressure $c_{Pp}$. The dispersed phase is further described by its volume concentration $\alpha(\bm{x},t)$, velocity $\textbf{w}(\bm{x},t)$, and temperature $T_p(\bm{x},t)$ fields. The governing conservation equations for mass, momentum, and heat in both the continuous and dispersed phases are written as follows: 
\begin{eqnarray}
      0 &=&  \nabla \cdot \textbf{u} \quad , \label{eq:mass-fluid}\\
d_t \alpha  &=& - \alpha \, (\nabla \cdot \textbf{w}) \quad ,\label{eq:mass-particles}\\
 D_t\mathbf{u}  &=& - \frac{1}{\rho}\,\nabla p +\nu\,\nabla^2 \mathbf{u}   + [\,1-\beta_T(T-T_r)\,]\,\mathbf{g} \\ \nonumber &+& \alpha \left[ \left(D_t\mathbf{u} - \mathbf{g}\right) + \frac{3-\beta}{2 \beta} \left(\mathbf{g} - d_t\mathbf{w}  \right) \right] \quad , \label{eq:mom-fluid}\\
d_t \mathbf{w} &=&  \beta D_t\mathbf{u}+\frac{12\,\nu\,\beta}{\textrm{d}_p^2}(\mathbf{u}-\mathbf{w})+(1-\beta)\mathbf{g} \quad ,\label{eq:mom-particle}\\
   D_t T   &=& \kappa\,\nabla^2  T  + \alpha \left[ {D_t T} - \frac{12 \kappa}{\textrm{d}_p^2}\,(T - T_p)\right]\quad \mbox{and}\label{eq:temp-fluid}\\
     d_t T_p&=& \frac{12 \kappa}{\textrm{d}_p^2}\,\frac{2\beta}{3-\beta}\,\frac{1}{\epsilon}\,(T-T_p).\label{eq:temp-particle}
\end{eqnarray}

A few additional observations are in order. The operators
$D_t() = \partial_t() + \textbf{u} \cdot \nabla()$
and
$d_t() = \partial_t() + \textbf{w} \cdot \nabla()$
denote the material derivatives for the fluid and particulate phases, respectively, where $\nabla()$ is the spatial gradient operator. Three hydrodynamic forces are considered: (i) Stokes drag, (ii) added-mass correction to fluid acceleration, and (iii) buoyancy. The drag force is characterized by the viscous response time
$\tau_p = \textrm{d}_p^2/(12\,\nu\,\beta)$,
where $\nu$ is the fluid kinematic viscosity. The added-mass contribution depends on the modified density ratio
$\beta = 3\rho/(\rho + 2\rho_p)$,
where $\rho$ is the fluid density. The particle temperature is assumed spatially uniform (lumped approximation), with relaxation toward the local fluid temperature governed by the timescale
$\tau_T = (3-\beta)/(2\beta)\,\epsilon\,\textrm{d}_p^2/(12 \kappa)$,
where $\kappa$ is the fluid thermal diffusivity and
$\epsilon = c_{Pp}/c_p$
is the ratio of particle to fluid specific heat capacity at constant pressure. The remaining constants include the fluid volumetric thermal expansion coefficient $\beta_T$ at the reference temperature $T_r$, the gravitational acceleration vector $\textbf{g}$, and the pressure field $p(\textbf{x},t)$. \textcolor{black}{We stress that the model takes into account thermal and mechanical couplings between the fluid and the particles (two-way) but neglects particle-particle (four-way) couplings, e.g. contact collisions and lubrication forces.} 

The domain considered here is three-dimensional, unbounded in the horizontal directions, and vertically confined between two parallel walls located at $z = \pm H/2$, with the unit vector $\hat{\textbf{z}}$ pointing upward. The fluid satisfies no-slip boundary conditions at both walls ($\textbf{u} = 0$), which are kept isothermal with a temperature difference $\Delta T$. The bottom wall is warmer, creating an unstable density stratification when $\beta_T > 0$. Particles are injected from one wall at a constant volumetric flux and prescribed velocity $\textbf{w}^{\ast}$, expressed as a multiple of the terminal velocity,
$\textbf{w}_T = (1-\beta)\,\tau_p\,\textbf{g}$.
Particles with $\beta < 1$, hereafter referred to as heavy particles, are injected from the top wall, whereas those with $\beta > 1$, referred to as light particles, are injected from the bottom wall. The particle inlet temperature is fixed at $T_p^{\ast}$, specified later. Particle accumulation at the opposite wall is neglected, and particles are assumed to leave the domain immediately upon arrival. Because the particulate-phase equations are first order in space and lack diffusive terms, only inlet boundary conditions are required to determine their solution. 

\subsection{Dimensionless system}

This model is rewritten in dimensionless form using the characteristic height $H$, the thermal diffusion timescale $H^2/\kappa$, and the fluid density $\rho$. The corresponding nondimensional variables are defined as
$\mathbf{U} = \mathbf{u}H/\kappa$,
$P = p H^2/(\rho \kappa^2)$,
$\Theta = (T - T_r)/\Delta T$,
$\mathbf{W} = \mathbf{w}H/\kappa$, and
$\Theta_p = (T_p - T_r)/\Delta T$,
representing the dimensionless velocity, pressure, and temperature fields of the fluid and dispersed phases. Using the same notation for the dimensionless derivatives and differential operators, and defining the dimensionless space and time as $\mathbf{X} = (X, Y, Z)$ and $\mathcal{T}$, the governing equations (\ref{eq:mass-fluid})–(\ref{eq:temp-particle}) can be rewritten as follows:
\begin{eqnarray}
0 &=& \nabla \cdot \textbf{U}\quad,\label{massfluidunperturb}\\
d_{\mathcal{T}} \alpha &=& - \alpha\,(\nabla \cdot \textbf{W})\quad,\label{massparticleunperturb}\\
D_{\mathcal{T}}\mathbf{U} &=& -\nabla P + Pr\,\nabla^2 \mathbf{U} + Pr\,Ra\,\Theta\,\hat{\mathbf{Z}} \nonumber \\
&+& \frac{\alpha}{2} \left[ (\beta-1) \left( D_{\mathcal{T}}\mathbf{U} + Ga\,Pr^{2}\,\hat{\mathbf{Z}} \right) - \frac{12\,Pr\,(3-\beta )}{\Phi^{2}}\left( \mathbf{U} - \mathbf{W} \right) \right] \,\,,
    \label{momentumfluidunperturb}\\
d_\mathcal{T} \mathbf{W}
 &=&  \beta \left( D_{\mathcal{T}}\mathbf{U} + \frac{12\,Pr}{\Phi^{2}} \left(\mathbf{U} - \mathbf{W}\right) \right) - (1 - \beta)\,Ga\,Pr^{2}\,\hat{\mathbf{Z}}\quad,
\label{momentumparticleunperturb}\\
D_{\mathcal{T}}\Theta &=& {\nabla^2 \Theta} + \alpha\left[{D_{\mathcal{T}}\Theta}- \frac{12}{\Phi^{2}}\left(\Theta - \Theta_p \right)\right] \quad \mbox{and}
\label{energyfluidunperturb}\\
d_{\mathcal{T}} \Theta_p &=& \frac{12}{\Phi^{2} \epsilon}\frac{2\,\beta}{3-\beta}\left(\Theta - \Theta_p \right) \quad ,
    \label{energyparticleunperturb}
\end{eqnarray}
subject to the boundary conditions
\begin{equation}
\mathbf{U} = 0\,\,\,\,,\,\,\,\, \Theta = 1 \,\,\,\,\text{at}\,\,\,\, Z = -1/2\,\,\,\,,\,\,\,\, \mathbf{U} = 0\,\,\,\,,\,,\, \Theta = 0 \,\,\,\,\text{at}\,\,\,\, Z = +1/2 \quad,
\label{eq:bc:fluid}
\end{equation}
\begin{equation}
\mathbf{W} = \mathbf{W^{\ast}} = W^{\ast}\hat{\mathbf{Z}}\,\,\,\,,\,\,\,\, \alpha = \mathcal{J}/||\mathbf{W^{\ast}}||\,\,\,\,,\,\,\,\, \Theta_p = \Theta_p^{\ast} \,\,\,\,\text{at}\,\,\,\, Z = Z^{\ast} \quad ,
\label{eq:bc:particle}
\end{equation}
where $Z^{\ast}$ is the position of the injection wall, which can be either $Z^{\ast}=+1/2$ or $-1/2$. The sign of $W^{\ast}$ depends on whether the particles are heavier or lighter than the fluid, while the inlet flux $\mathcal{J}$ is defined as positive.

In the dimensionless formulation, the characteristic parameters are defined as
\begin{equation}
Ra=\frac{\beta_T \Delta T g H^3}{\nu \kappa} \quad,\quad Pr = \frac{\nu }{\kappa} \quad,\quad Ga = \frac{g H^3}{\nu^2} \quad\mbox{and}\quad \Phi = \frac{\textrm{d}_p}{H} \quad, \label{eq:control-param}
\end{equation}
where $Ra$ is the Rayleigh number, representing the ratio of buoyant to diffusive effects, $Pr$ is the Prandtl number, expressing the relative importance of momentum and thermal diffusivities, $Ga$ is the Galileo number, quantifying the balance between gravitational and viscous forces, and $\Phi$ is the particle-to-domain size ratio. Although $Ga$ does not depend on particle properties, it becomes a relevant control parameter when particle–fluid coupling is included. In dimensionless form, the terminal particle velocity is
$\mathbf{W}_T = (\beta -1)/\beta\,(\Phi^2/12)\,Ga\,Pr\,\hat{\mathbf{Z}}$
. Together with the modified density ratio $\beta$, the particle inlet flux $\mathcal{J}$, and the inlet velocity and temperature $(W^{\ast}, \Theta_p^{\ast})$, these quantities define the full set of control parameters. In total, the model is governed by nine parameters: three associated with the fluid phase $(Ra, Pr, Ga)$ and six associated with the particulate phase $(\Phi, \beta, \epsilon, \mathcal{J}, W^{\ast}, \Theta_p^{\ast})$.  

\section{\label{sec:stability}Linear stability analysis}

\subsection{Conductive state}
  
To determine the onset of natural convection in the pRB system, its conductive equilibrium state must first be defined. This steady-state corresponds to a quiescent configuration where particles either settle or rise in a purely conductive fluid. Accordingly, one imposes $\mathbf{U}=0$, $\mathbf{W}=W_0(Z)\,\hat{\mathbf{Z}}$, $P=P_0(Z)$, $\alpha=\alpha_0(Z)$, $\Theta=\Theta_0(Z)$, and $\Theta_p=\Theta_{p0}(Z)$, and rewrites equations (\ref{massfluidunperturb})–(\ref{energyparticleunperturb}) as:
\begin{equation}
    D(\alpha_0 W_0) = 0 \quad ,
    \label{partice-mass-cons}
\end{equation}
\begin{equation}
0 = -DP_0 + Pr\,Ra\,\Theta_0\,\hat{\mathbf{Z}}
+ \frac{\alpha_0}{2}\,(\beta-1)\,Ga\,Pr^{2}\,\hat{\mathbf{Z}} + \frac{6\,\alpha_0\,Pr\,(3-\beta )}{\Phi^{2}}\,W_0 \quad ,
\end{equation}
\begin{equation}
    W_0\,DW_0 = -\frac{12\,Pr\,\beta\,W_0}{\Phi^{2}} - (1-\beta)\,Ga\,Pr^{2}\, \hat{\mathbf{Z}} \quad ,
\end{equation}
\begin{equation}
     D^2 \Theta_0 - \frac{12\,\alpha_0}{\Phi^{2}}\,(\Theta_0-\Theta_{p0}) = 0 \quad \mbox{and}
     \label{eq16}
\end{equation}
\begin{equation}
 W_0 D\Theta_{p0} =\frac{12}{\Phi^{2}\,\epsilon} \, \frac{2\,\beta}{3-\beta} \, (\Theta_0-\Theta_{p0}) \quad ,
\label{eq15}
\end{equation}
where $D$ denotes differentiation with respect to $Z$. The steady-state equations (\ref{partice-mass-cons})–(\ref{eq15}), together with boundary conditions derived from (\ref{eq:bc:fluid}) and (\ref{eq:bc:particle}), are solved numerically to obtain the base-state fields. 

\begin{figure}[!htb]
     \centering
    \begin{subfigure}[c]{0.49\textwidth}
         \centering
           \subcaption{$\beta = 0.5$ (Heavy particles)}\includegraphics[width=\textwidth]{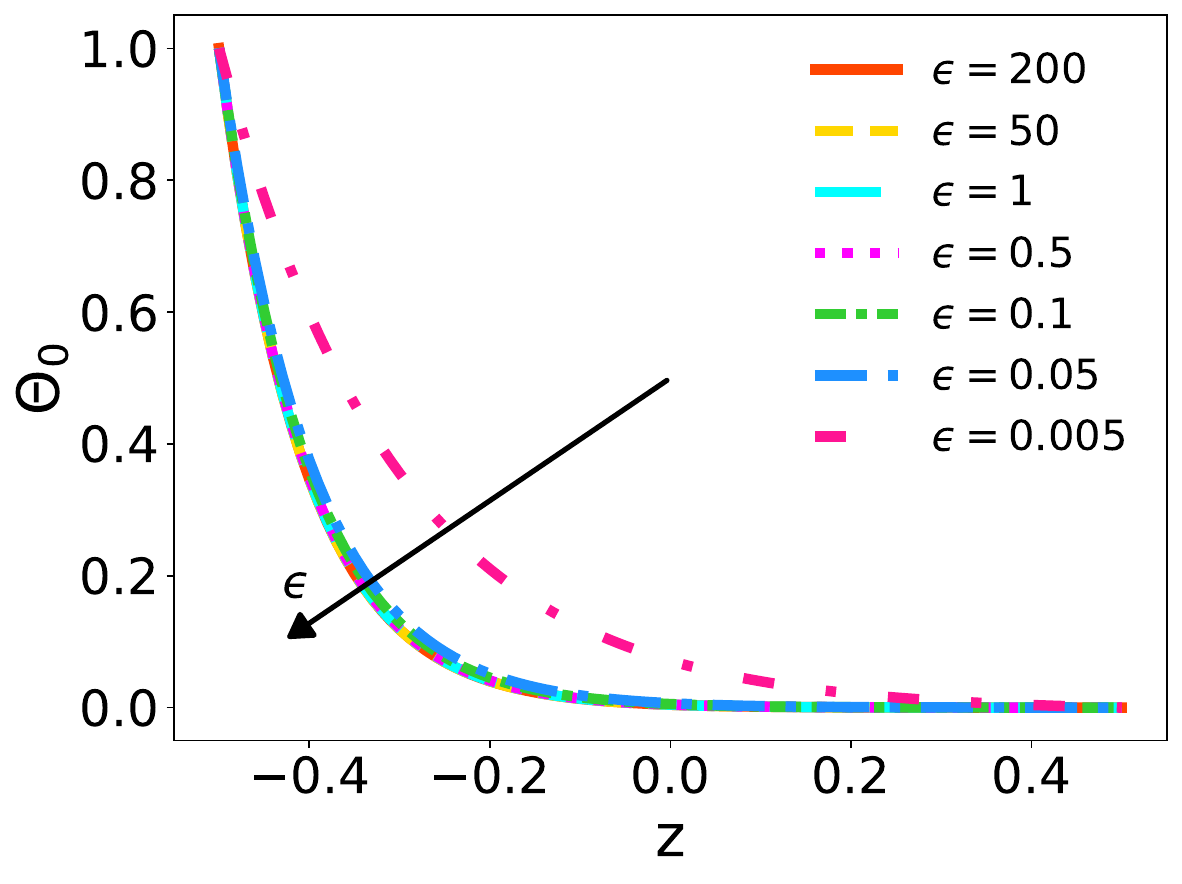}
       
         \end{subfigure}
         \hfill
         \begin{subfigure}[c]{0.49\textwidth}
         \centering
            \subcaption{$\beta = 0.5$ (Heavy particles)}\includegraphics[width=\textwidth]{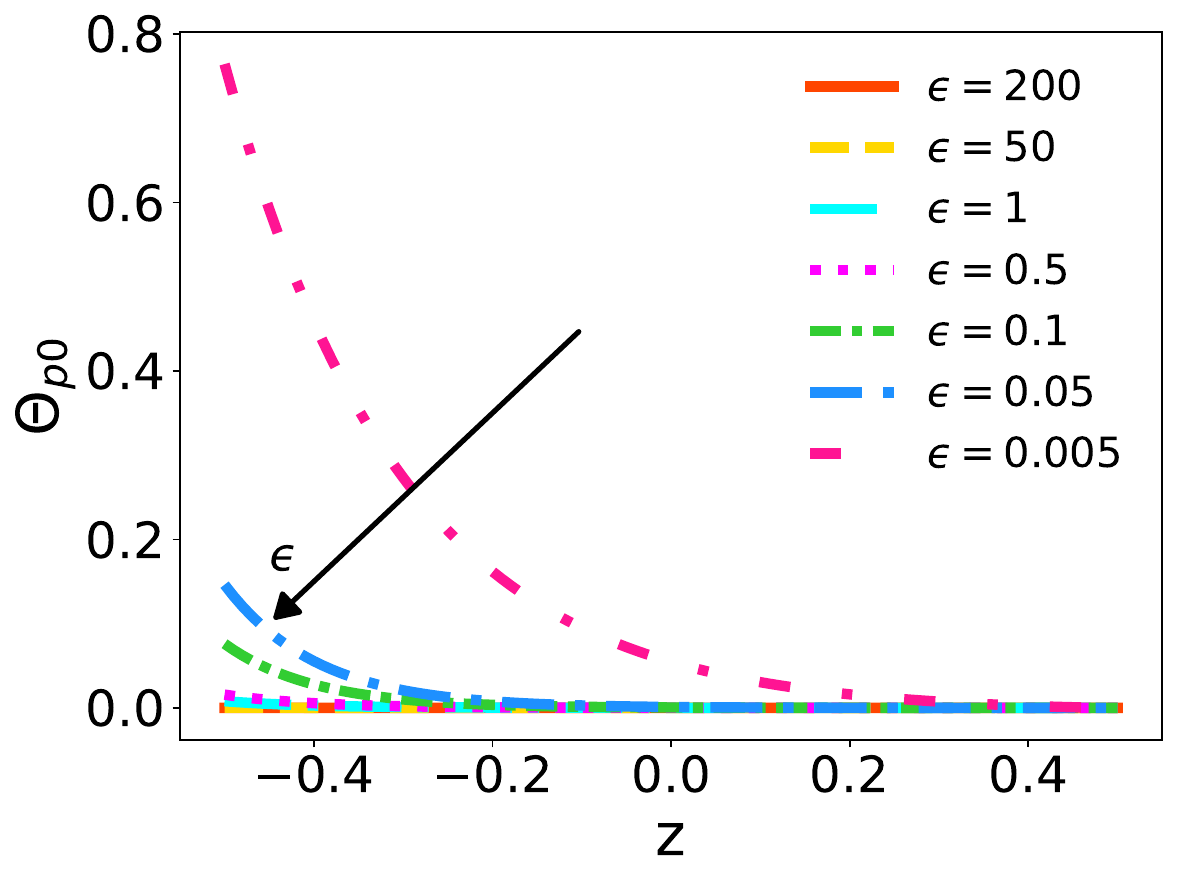}
      
         \end{subfigure}
     \begin{subfigure}[c]{0.49\textwidth}
         \centering
          \subcaption{$\beta = 2.5$ (Light particles)}\includegraphics[width=\textwidth]{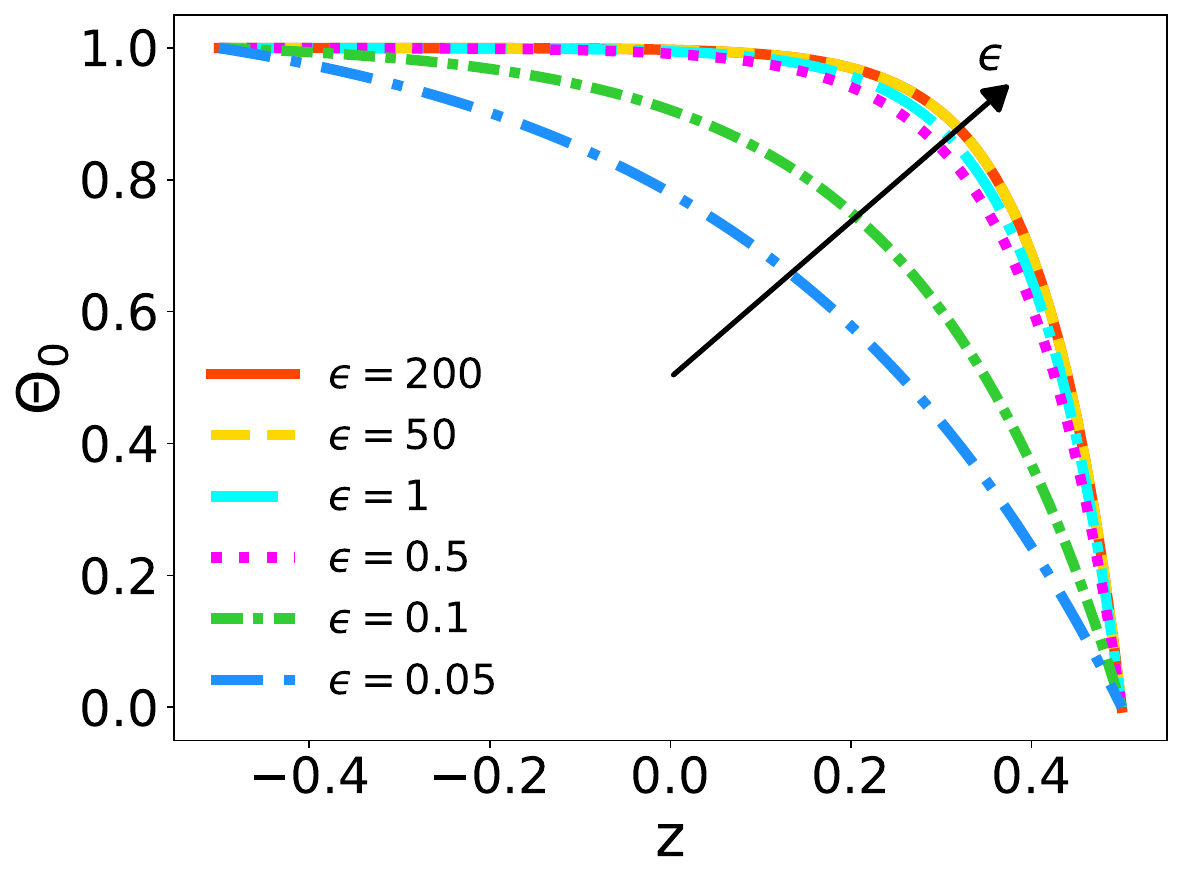}
        
         \end{subfigure}
         \hfill
         \begin{subfigure}[c]{0.49\textwidth}
         \centering
           \subcaption{$\beta = 2.5$ (Light particles)}\includegraphics[width=\textwidth]{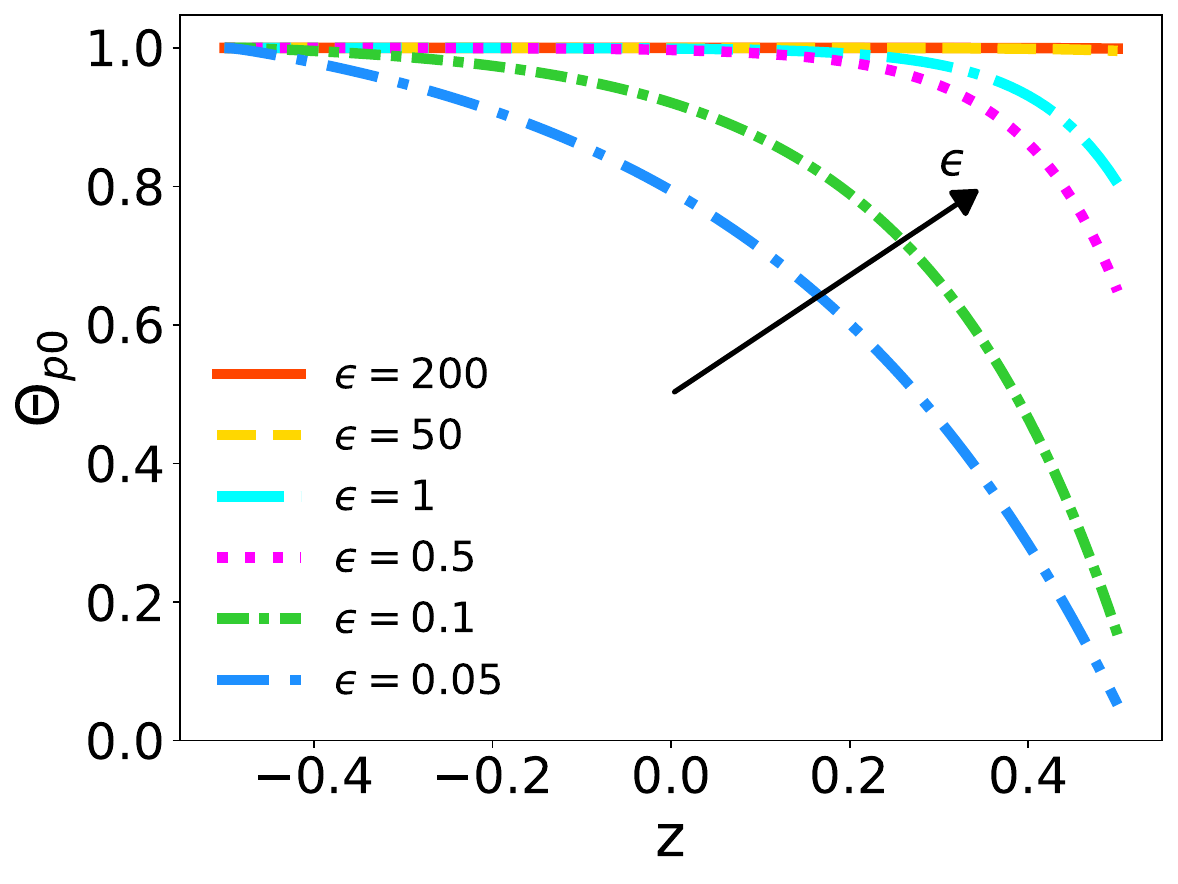}
         \end{subfigure}

  \caption{Vertical dependence of the steady-state temperature fields of the (left) fluid $\Theta_0(Z)$ and (right) particulate $\Theta_{p0}(Z)$ phases, for (bottom) light and (top) heavy particles, and different values of the specific heat capacity ratio $\epsilon$. The particle flux is set to $\mathcal{J}_0= 533.3$, $\mathbf{W}^{\ast} = 0.5\,\mathbf{W}_T$, and the particle injection temperature $\Theta_p^{\ast}$ is assumed to match the temperature of the wall from which they are introduced. The arrow indicate the direction of increase of $\epsilon$.}
    \label{fig:base-state}
\end{figure}

The fluid and particle temperature base profiles are shown in Figure \ref{fig:base-state} for different values of the specific heat capacity ratio $\epsilon$. This parameter sets the strength of thermal coupling: small $\epsilon$ yields rapid particle–fluid thermal equilibration, whereas large $\epsilon$ implies a slower particle response, effectively acting as distributed heat sources or sinks. Consequently, increasing $\epsilon$ homogenizes the fluid temperature over most of the domain and confines the imposed gradient to a narrow region near the particle extraction wall. In general, particle concentration varies with height, except when particles are injected at their terminal velocity ($W^{\ast}=W_T$), which yields a uniform concentration~\citep{prakhar2021linear,razaPF2024}. For $W^{\ast}<W_T$ ($W^{\ast}>W_T$), particles accelerate (decelerate) toward $W_T$, producing accumulation (rarefaction) near the injection wall -- upper region for heavy particles and lower region for light ones. For the prescribed inflow used here, reducing $W^{\ast}$ increases the inlet volumetric concentration ($\alpha=\mathcal{J}/||\mathbf{W}^{\ast}||$), i.e. smaller inlet velocities lead to higher particle loading. Finally, the pressure base state $P_0$ has no dynamical role in the analysis that follows. 

\subsection{Perturbation equations}

Having defined the equilibrium state of the pRB system, the next step is to examine its linear stability. This is achieved by decomposing all variables into their steady-state and small-amplitude perturbations, linearizing the resulting equations, and assuming the perturbations can be represented as normal modes of the form
\begin{equation}
\xi'(X,Z,\mathcal{T})=\xi^n(Z) \exp(i k X+\lambda \mathcal{T}) + \textit{c.c.}\label{eq:normalmodes}
\end{equation}
where $\xi' = {\mathbf{U}', \mathbf{W}', \alpha', \Theta', \Theta_p'}$ is the vector of perturbed quantities, \textit{c.c.} denotes the complex conjugate, and $\xi^n(Z)$ is the normal-mode amplitude varying along the non-homogeneous direction $Z$. In the temporal stability framework, $k$ is the real wavenumber, and $\lambda = \lambda_r + i\lambda_i$, with $\lambda_r$ representing the temporal growth rate of the perturbation and $\lambda_i$ its oscillation frequency. Substituting these definitions into the governing equations yields the following linear system for the perturbed quantities:
\begin{equation}
     \lambda \alpha^n  = -\big(\alpha_0 DW^n_{z} + W^n_{z}{D\alpha_0}\big)  - \alpha_0 \Dot{\iota} k W^n_{{x}} - \big(W_0 D\alpha^n  + \alpha^n {D W_0}\big) \quad ,
     \label{eq30}
\end{equation}
\begin{eqnarray}
\lambda \,(D^2 -k^2) {U^n_z} \!\!\!&=&\!\!\!  Pr\,(D^2 - k^2)^2\,U^n_z - Pr\,Ra\,k^2 
\Theta^n \nonumber \\
&+&\!\!\! \frac{\alpha_0}{2}\,(\beta -1)\,\lambda\,(D^2 -k^2)\,U_z^n \nonumber \\
&+&\!\!\! {\frac{D\alpha_0 }{2}\,(\beta -1)\,\lambda D U_z^n - \frac{D\alpha_0\,6\,Pr\,(3 -\beta)}{\Phi^{2}} \, (DU_z^n  + \dot\iota\,k\,W^n_x )} \nonumber \\
&-&\!\!\! \dfrac{\alpha_0\,6\,Pr\,(3 -\beta)}{\Phi^{2}}(\dot\iota\,k\,D\,W_x^n + (D^2-k^2)\,U_z^n +k^2\,W_z^n) \nonumber \\
&-&\!\!\!\Bigg[\frac{1}{2}\,(\beta -1)\,Ga\,Pr^{2} + \dfrac{6\,Pr\,(3 -\beta)\,W_0}{\Phi^{2}} \Bigg] \, k^2\,\alpha^n \quad ,
\label{eq35}
\end{eqnarray}
\begin{equation}
 \lambda {W_z}^n + W_0 {D {W_z}^n} +  {W_z}^n {DW_0} = \beta \lambda {U^n_z} + \dfrac{12 Pr \beta }{\Phi^{2}}{({U_z}^n-{W_z}^n)} \quad ,
 \label{eq37}
\end{equation}
\begin{equation}
 \lambda {W_x}^n + W_0 ({D{W_x}^n}) = \beta \lambda {U^n_x} + \dfrac{12 Pr \beta }{\Phi^{2}}{({U_x}^n-{W_x}^n)} \quad ,
 \label{eq37}
\end{equation}
\begin{eqnarray}
{(1-\alpha_0) } \left[\lambda {\Theta^n} + U^n_z D \Theta_0\right] \!\!\!&=&\!\!\! \alpha^n \lambda \Theta_0 + (D^2 - k^2) \Theta^n   - \dfrac{12 \alpha_0}  {\Phi^{2}}{(\Theta^n - \Theta_{p}^n)} \nonumber \\
&-&\!\!\!  \dfrac{12 \alpha^n}{\Phi^{2}}{(\Theta_0 - \Theta_{p0})} \quad \mbox{and}
 \label{eq38}
\end{eqnarray}
\begin{equation}
     \lambda  \Theta^n_p + W_0 D\Theta^n_p + {W_z}^n  D \Theta_{p0} = \frac{12}{\Phi^{2} \epsilon}\frac{2 \beta}{3-\beta}{(\Theta^n - \Theta^n_p)} \quad ,
     \label{eq39}
\end{equation}
obtained by taking the double curl of the fluid momentum equation to eliminate the pressure term and by applying the incompressibility condition to remove the $x$-component of the velocity. The corresponding boundary conditions, derived in the same way from (\ref{eq:bc:fluid}) and (\ref{eq:bc:particle}), are
\begin{eqnarray}
\alpha^n=\bm{W^n} = U_z^n = DU_z^n = \Theta^n = 0 & \text{at} & Z = \pm{1}/{2} \quad \mbox{and} \nonumber \\
\Theta^n_{p} = 0 & \text{at} & Z = Z^{\ast} \quad . \label{eq40}
\end{eqnarray}

Equations (\ref{eq30})–(\ref{eq40}) are solved numerically using the shooting method, \textcolor{black}{with the critical conditions ($Ra_c, \lambda_c, k_c$) identified by minimizing $Ra$ with respect to $k$}~\citep{Alvesetal}. To verify the results, a matrix-forming approach is also employed. In this formulation, the differential system is discretized with a fourth-order finite-difference scheme and cast as a generalized algebraic eigenvalue problem. The resulting matrix system is then solved using the Arnoldi method with a shift-and-invert spectral transformation~\citep{Souzaetal}. Both methods show excellent quantitative agreement, as shown in Table~\ref{tab:comparison_Ra_kc}.

\section{\label{sec:results}Results and discussion}

In this study, some parameters are kept constant: $Pr = 5$, $Ga = 1.92\times10^{10}\ $\footnote{It is worth noting that this value matches the one used in our previous works: \citep{Raza_Freitas_Alves_Calzavarini_Hirata_2025}, where it was inadvertently reported incorrectly, and \citep{razaPF2024}, where it appeared within the parameter $\Lambda = Ga Pr^2$, not used in the current study.}, and $\Phi = 10^{-2}$. The critical Rayleigh number $Ra_c$ serves as the primary control variable for the stability analysis. The parameters varied are $\beta$, $\epsilon$, $\Theta_p^{\ast}$, $\mathcal{J}$, and $\mathbf{W}^{\ast}$. Unless otherwise stated, the particle flux and injection velocity are set to $\mathcal{J}_0 = 533.3$ and $\mathbf{W}^{\ast} = 0.5\mathbf{W}_T$, respectively, whereas the particle injection temperature $\Theta_p^{\ast}$ equals the temperature of the wall from which the particles are introduced—hot ($\Theta_p^{\ast} = 1$) for light particles injected from below, and cold ($\Theta_p^{\ast} = 0$) for heavy particles injected from above. 

\subsection{Influence of the specific heat capacity ratio}

\begin{figure}[!htb]
    \centering
    \begin{minipage}{0.76\linewidth}
        \centering
            \includegraphics[width=\textwidth]{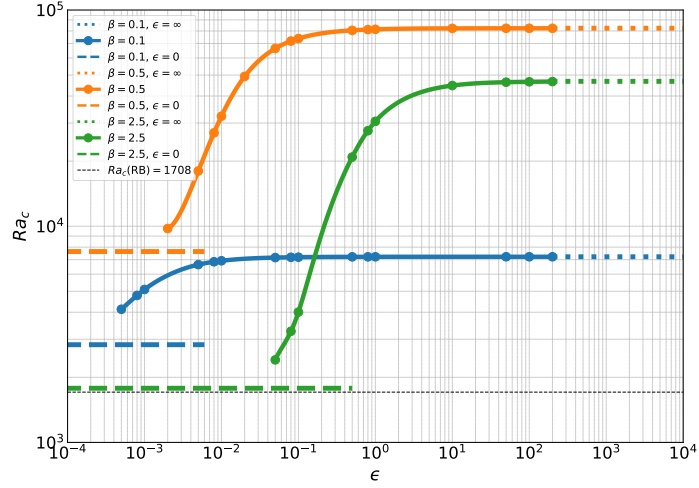}
         \subcaption{}
         \label{fig1Ra}
    \end{minipage}

    \vspace{1.5em}  

    \begin{minipage}{0.74\linewidth}
        \centering
       \includegraphics[width=\textwidth]{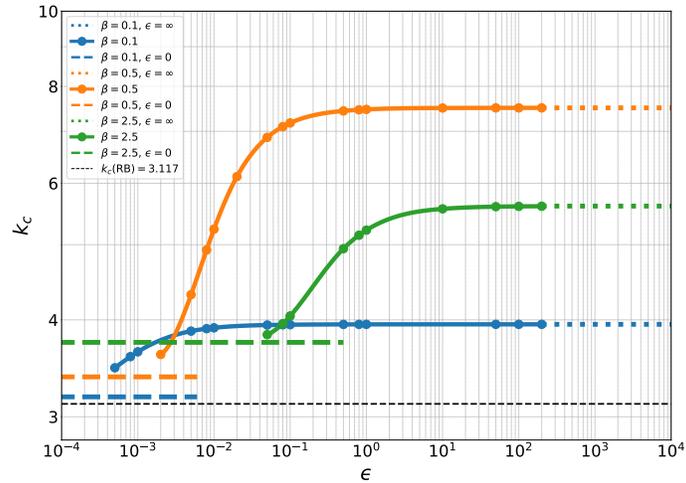}
         \subcaption{}
         \label{fig1kc}
    \end{minipage}
 \caption{Effect of thermal coupling on the critical stability thresholds of the particulate Rayleigh-Bénard (pRB) system for heavy and light particles. Panel (a) shows the critical Rayleigh number $Ra_c$, while panel (b) presents the critical wavenumber $k_c$. The horizontal dashed colored lines correspond to the limiting case $\epsilon = 0$, which represents purely mechanical coupling without any thermal effects, while the dotted horizontal lines represents the limiting case  $\epsilon \rightarrow \infty$ , where the instability reaches a clear plateau. The black dashed line indicates the reference Rayleigh–Bénard threshold for a single-phase system $Ra_c=1708$.}
    \label{fig:fig1}
\end{figure}

Figure \ref{fig:fig1} shows the dependence of the critical Rayleigh number $Ra_c$ on the specific heat capacity ratio $\epsilon$ for representative values of $\beta$ corresponding to heavy and light particles. The parameter $\epsilon$ quantifies the efficiency of heat exchange between particles and fluid, thereby controlling the strength of thermal coupling. For small $\epsilon$, the particle heat capacity is negligible compared to that of the fluid, and the particles rapidly equilibrate with the surrounding temperature field. In contrast, for large $\epsilon$, particles act as distributed heat sources or sinks, influencing the local thermal field while maintaining their own temperature. \textcolor{black}{While the limit $\epsilon \rightarrow \infty$ is idealized, large but finite values of $\epsilon$ can be physically relevant. For example, the specific heat capacity ratio of ice particles in air is $\sim2$ while cryogenic liquid hydrogen drops in mercury (Hg) it is $\sim 100$. In contrast, for solid particles such as sand in water ratio becomes $\sim 0.19$ and for metal particles such as cooper (Cu) in water the ratio is $\sim0.09$.} The mechanical coupling between the two phases remains unaffected, as it is independent of $\epsilon$. This separation of mechanical and thermal effects distinguishes the present parametrization from earlier pRB studies.

The results in Figure \ref{fig:fig1} show that stronger thermal coupling systematically increases system stability, as reflected by higher $Ra_c$ values for both heavy and light particles. The stabilizing trend persists until a saturation point, beyond which further increases in $\epsilon$ no longer affect the critical thresholds. This saturation behavior is also observed in the critical wavenumber $k_c$. In case of extreme heat capacity ratio ($\epsilon \rightarrow \infty$) the temperature of the particulate phase does not change and also in this case particle energy equation can be discarded. We have to reconstruct the base state which follows from equations (\ref{energyfluidunperturb}) and (\ref{energyparticleunperturb}) as,

\begin{equation}
    D^2 \Theta_0 - \frac{12 \alpha_0}{\Phi^2} (\Theta_0 - \Theta_p^*) = 0 
    \label{base_fluid}
\end{equation}
 
Finally, the stabilizing effect is not monotonic with $\beta$ for the three representative cases shown in Figure \ref{fig:fig1}; this dependence is further discussed in Section~\ref{subsec:mass_density}.

\subsection{Impact of injection temperature}

\begin{figure}[!htb]
     \centering
     \begin{subfigure}[c]{0.49\textwidth}
         \centering
         \includegraphics[width=\textwidth]{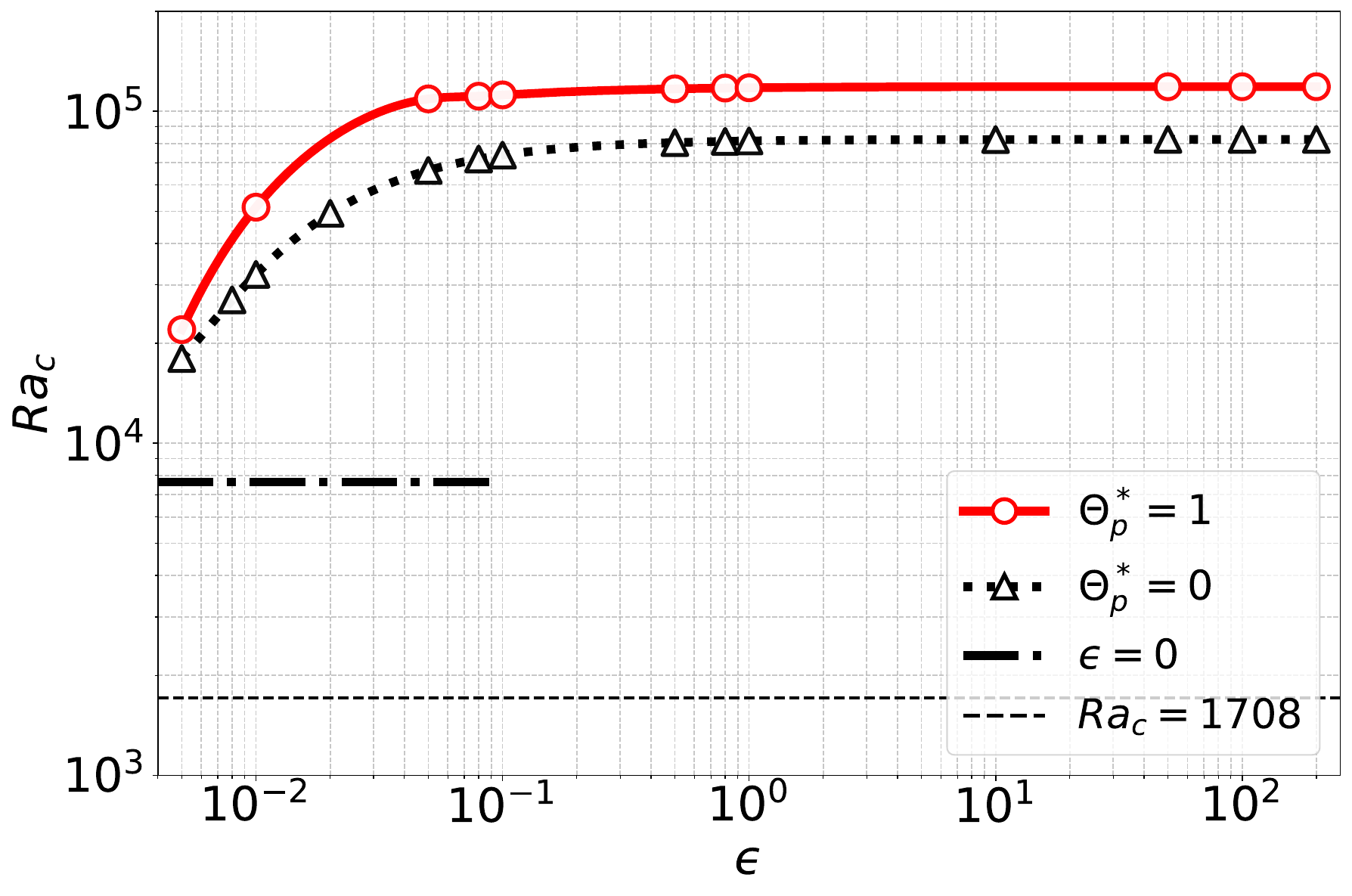}
         \subcaption{$\beta = 0.5$ (Heavy particles)}
         \end{subfigure}
         \hfill
         \begin{subfigure}[c]{0.49\textwidth}
         \centering
         \includegraphics[width=\textwidth]{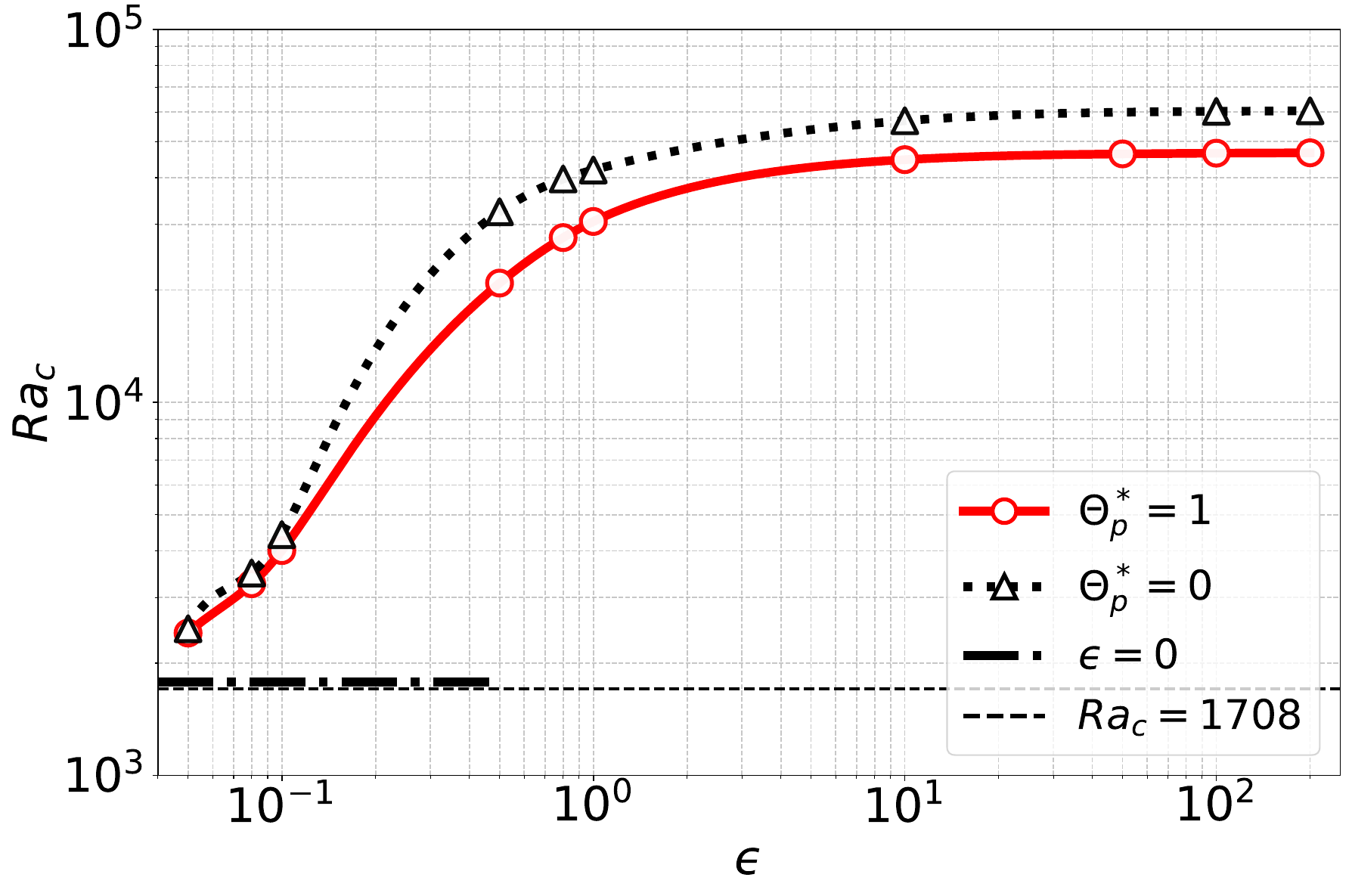}
         \subcaption{$\beta = 2.5$ (Light particles)}
         \end{subfigure}
     \caption{Effect of particle injection temperature on the critical threshold. (a) Heavy particles ($\beta = 0.5$) are injected from above with the cold wall temperature $\Theta_p^* = 0$, and inversely with the temperature of the opposite hot wall, $\Theta_p^* = 1$.  (b) Light particles ($\beta = 2.5$) are injected from below with the hot wall temperature $\Theta_p^* = 1$, and inversely with the opposite cold wall temperature $\Theta_p^* = 0$.}
        \label{fig:fig2}
\end{figure}

Figure \ref{fig:fig2} shows that the stabilizing effect of $\epsilon$ is independent of the particle injection temperature. This is evident from the similar trends of $Ra_c(\epsilon)$ for cases where particles enter the system with the temperature of the injection wall or with that of the opposite wall. Although the absolute $Ra_c$ values differ between these cases, the stabilizing tendency with increasing $\epsilon$ remains unchanged for both heavy and light particles. As $\epsilon \to 0$, the influence of the injection temperature vanishes, and both curves converge to the critical Rayleigh number of the purely mechanically coupled system.

\subsection{Influence of the particulate mass density}
\label{subsec:mass_density}

\begin{figure}[!htb]
    \centering
    \begin{minipage}{0.77\linewidth}
        \centering
        \includegraphics[width=\linewidth]{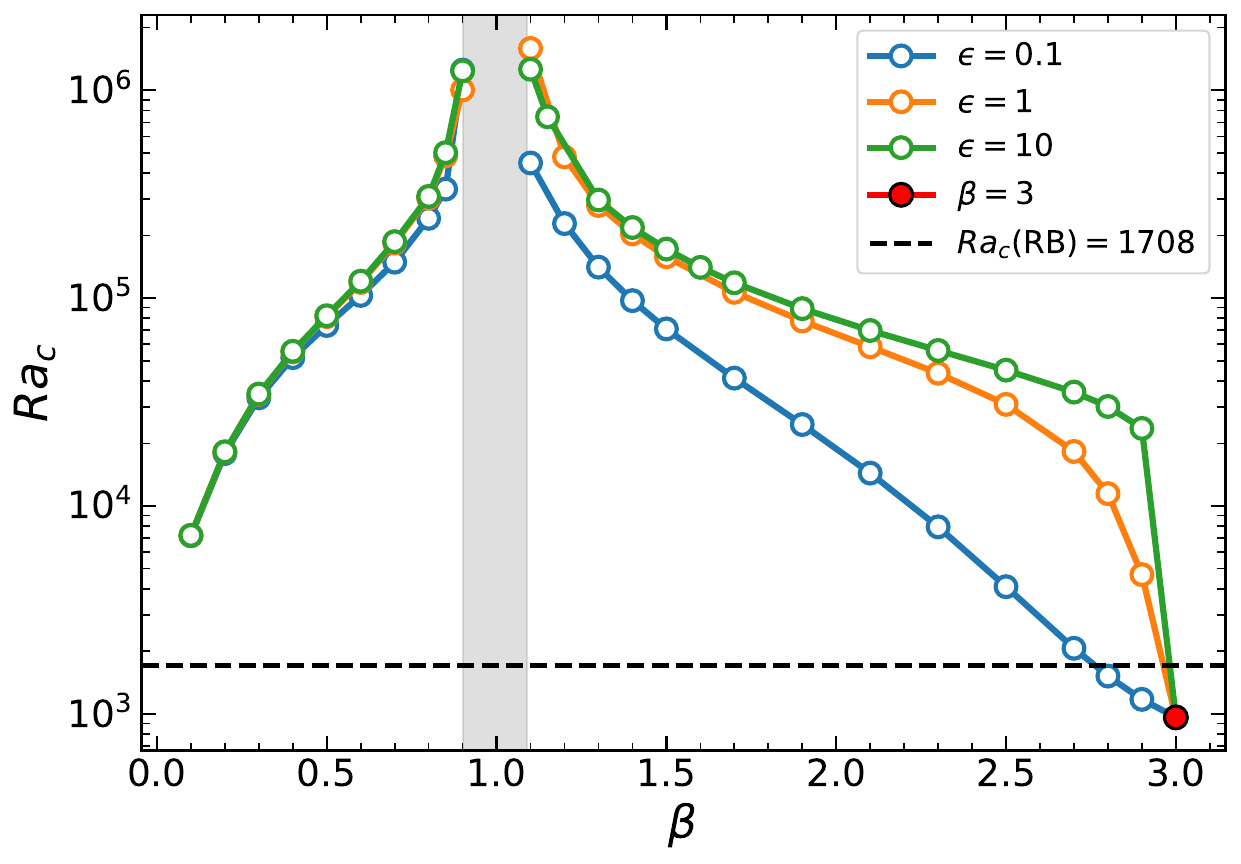}
        \subcaption{}
        \label{fig:RavBeta}
    \end{minipage}

    \vspace{1.5em}  

    \begin{minipage}{0.76\linewidth}
        \centering
        \includegraphics[width=\linewidth]{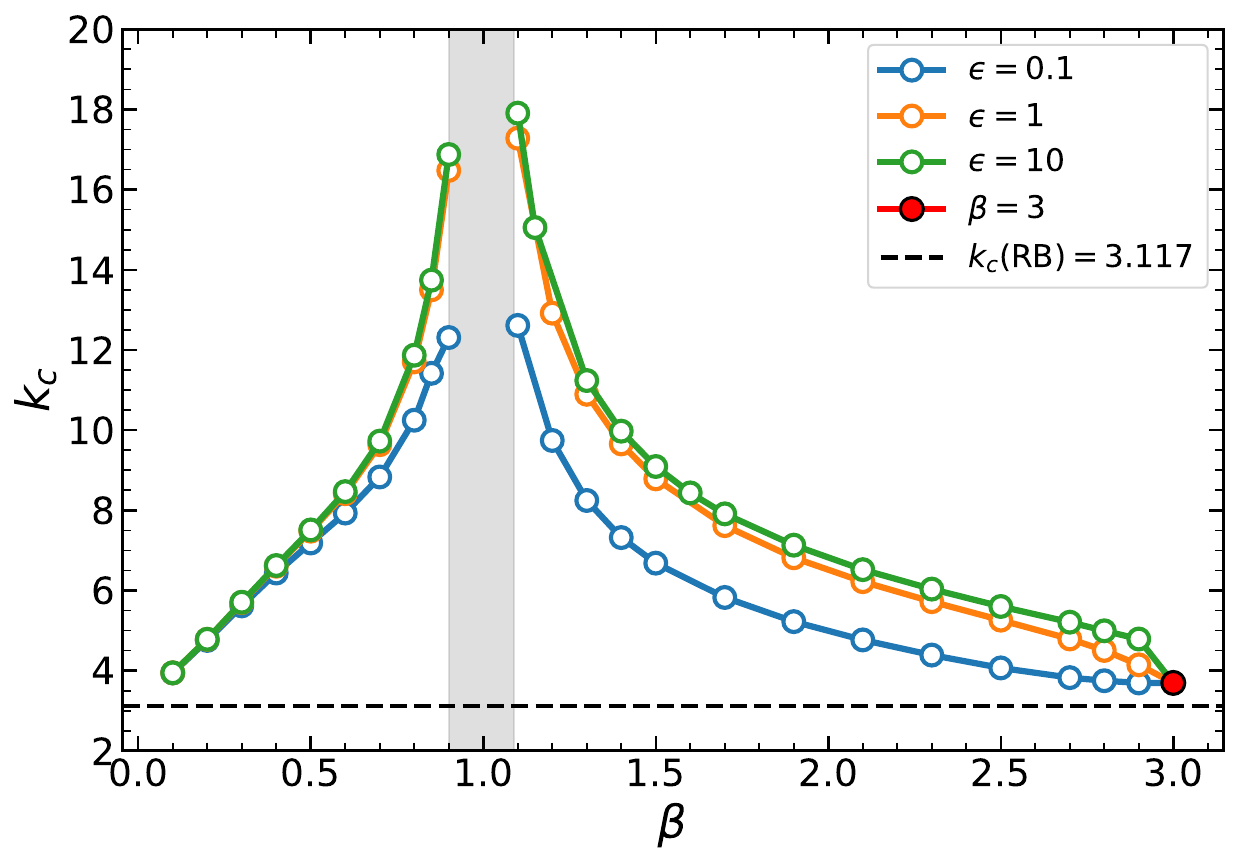}
        \subcaption{}
         \label{fig:KvBeta}
    \end{minipage}

    \caption{Impact of particle density on system stability for a wide range of density and three representative heat capacity ratios $\epsilon$. Panel (a) shows the critical Rayleigh number $Ra_c$ as a function of the modified density ratio $\beta$, with the black dashed line showing the reference Rayleigh–Bénard threshold corresponding to a single-phase system (\( Ra_c = 1708 \)). Panel (b), on the other hand, presents the critical wavenumber $k_c$ as a function of $\beta$, the dashed line represents the critical wavenumber in the single-phase case.The gray area in both graphs denotes the region where the model reaches its limit of validity. Specifically, the case $\beta = 1$ corresponds to a singular condition in which obtaining consistent results becomes difficult.}
    \label{fig:Rac_beta}
\end{figure}

Figures~\ref{fig:RavBeta} and~\ref{fig:KvBeta} show the respective dependence of the critical Rayleigh and wavenumbers, $Ra_c$ and $k_c$, on the density ratio parameter $\beta$ for three different values of the specific heat capacity ratio $\epsilon$. The horizontal dashed lines mark the corresponding reference Rayleigh–Bénard thresholds, $Ra_c = 1708$ and $k_c = 3.117$. Three distinct regimes can be identified.

On the left-hand side of both figures ($\beta < 1$), corresponding to particles denser than the fluid, the three curves nearly collapse onto one another, indicating that thermal coupling has little influence in this regime. Even in the extremely heavy particle limit ($\beta \to 0$), where particles behave as distributed constant-temperature sources, as suggested by Equations~(\ref{eq:temp-fluid}) and (\ref{eq:temp-particle}), the stability is dominated by mechanical coupling. 
For $\beta = 0$ (the ballistic limit), particles retain a constant temperature and accelerate uniformly, resulting in an infinite terminal velocity. Since in our analysis the particles are injected at a velocity that is a fraction of their terminal velocity and the particle flux is held constant,  this effectively means no particles can be injected. In this limit, the system thus reduces to the classical single-phase Rayleigh–Bénard problem, with a convective onset at $Ra_c \simeq 1708$.

As the particle density approaches that of the carrier ($\beta \to 1$), the system becomes increasingly stable, i.e. $Ra_c$ rises sharply, and the dominant disturbances become smaller, i.e. $k_c$ increases. The limit $\beta \to 1$ is singular in the present model: maintaining a constant inlet particulate flux would require a particle concentration exceeding the dilute limit, leading to nonphysical or divergent solutions. This likely explains why the two numerical methods no longer yield consistent results near $\beta \sim 1$, marked as the gray-shaded region in both figures. Only data points for which consistent results were obtained across all methods are shown.

In the light-particle regime ($\beta > 1$), distinct trends emerge. As $\beta$ increases beyond unity, the three curves separate, showing that the influence of the specific heat capacity ratio $\epsilon$ becomes significant. Consistent with previous results, larger $\epsilon$ values, i.e. stronger particle heat capacity relative to the fluid—lead to higher critical Rayleigh and wavenumbers, indicating increased stability and smaller dominant structures, respectively. When $\beta$ is sufficiently large, however, all curves shift downward. For instance, when $\epsilon = 0.1$, $Ra_c$ falls below the single-phase Rayleigh–Bénard threshold. Varying $\beta$ therefore modifies both mechanical and thermal coupling effects. This dual influence can be seen directly in Equation~(\ref{eq:temp-particle}), where the limit $\beta \to 3$ is equivalent to $\epsilon \to 0$. 
\textcolor{black}{
In this limit, the particles instantaneously adjust their temperature to that of the surrounding fluid. The resulting dimensional equation governing the fluid temperature is
\begin{equation}
D_t T = \frac{\kappa}{1 - \alpha},\nabla^2 T .
\end{equation}
This result indicates that bubbles ($\beta = 3$) and particles with vanishing specific heat capacity ($\epsilon = 0$) modify the thermal properties of the fluid solely through an excluded-volume effect. In particular, the effective thermal diffusivity of the fluid is enhanced by a factor $1/(1 - \alpha)$. In the present study, this correction is negligible, since the mean particle concentration is $O(10^{-3})$.}

\begin{table}[!htb]
\centering
\caption{Numerical results for $Ra_c$ and $k_c$ for a selected subset of particle classes with $\beta$ and $\epsilon$ values, corresponding to results in  Figure~(\ref{fig:Rac_beta}).  We compare results from Shooting method (SM) and Matrix-Forming methods (MFM).  The case $\beta=3$ is independent of the $\epsilon$ value, i.e., it corresponds to the case of no thermal coupling.}
\renewcommand{\arraystretch}{1.2}
\resizebox{\textwidth}{!}{%
\begin{tabular}{
c| 
S[table-format=1.1]
S[table-format=3.4e1]
S[table-format=2.4]
S[table-format=3.4e1]
S[table-format=2.4]
}
\hline
\hline
{$\epsilon$} & {$\beta$} & {$Ra_c$ (SM)} & {$k_c$ (SM)} & {$Ra_c$ (MFM)} & {$k_c$ (MFM)} \\
\hline
\hline
\multirow{6}{*}{0.1}
 & 0.1 & 7.1900e3 & 3.9431 & 7.1906e3 & 3.9430 \\
 & 0.5 & 7.3801e4 & 7.1761 & 7.3875e4 & 7.1856 \\
 & 0.8 & 2.4175e5 & 10.2235 & 2.4027e5 & 10.2477 \\
 & 2.5 & 4.0072e3 & 4.0464 & 4.0905e3 & 4.0687 \\
 & 2.7 & 2.0326e3 & 3.8158 & 2.0679e3 & 3.8267 \\
 & 2.8 & 1.5002e3 & 3.7425 & 1.5204e3 & 3.7493 \\
\hline
\multirow{6}{*}{1.0}
 & 0.1 & 7.2184e3 & 3.9469 & 7.2189e3 & 3.9468 \\
 & 0.5 & 8.1409e4 & 7.4711 & 8.1479e4 & 7.4799 \\
 & 0.8 & 3.0225e5 & 11.6887 & 3.0249e5 & 11.7043 \\
 & 2.5 & 3.0544e4 & 5.2199 & 3.0827e4 & 5.2603 \\
 & 2.7 & 1.8142e4 & 4.7702 & 1.8306e4 & 4.7913 \\
 & 2.8 & 1.1465e4 & 4.5051 & 1.1465e4 & 4.5050 \\
\hline
\multirow{6}{*}{10}
 & 0.1 & 7.2645e3 & 3.9477 & 7.2208e3 & 3.9470 \\
 & 0.5 & 8.2220e4 & 7.5019 & 8.2291e4 & 7.5106 \\
 & 0.8 & 3.0814e5 & 11.8536 & 3.0836e5 & 11.8685 \\
 & 2.5 & 4.4727e4 & 5.5611 & 4.5033e4 & 5.6012 \\
 & 2.7 & 3.5091e4 & 5.1823 & 3.5318e4 & 5.2096 \\
 & 2.8 & 3.0113e4 & 4.9950 & 3.0113e4 & 4.9950 \\
\hline
$\forall \epsilon \in [0,\infty)$ & 3.0 & 9.6150e2 & 3.6948 & 9.6621e2  &  3.7146\\
\hline
\hline
\end{tabular}%
}
\label{tab:comparison_Ra_kc}
\end{table}

\subsection{Impact of particle injection velocity and inlet flux}

\begin{figure}[!htb]
    \centering
    \begin{subfigure}[t]{0.48\textwidth}
        \centering
        \includegraphics[width=\linewidth]{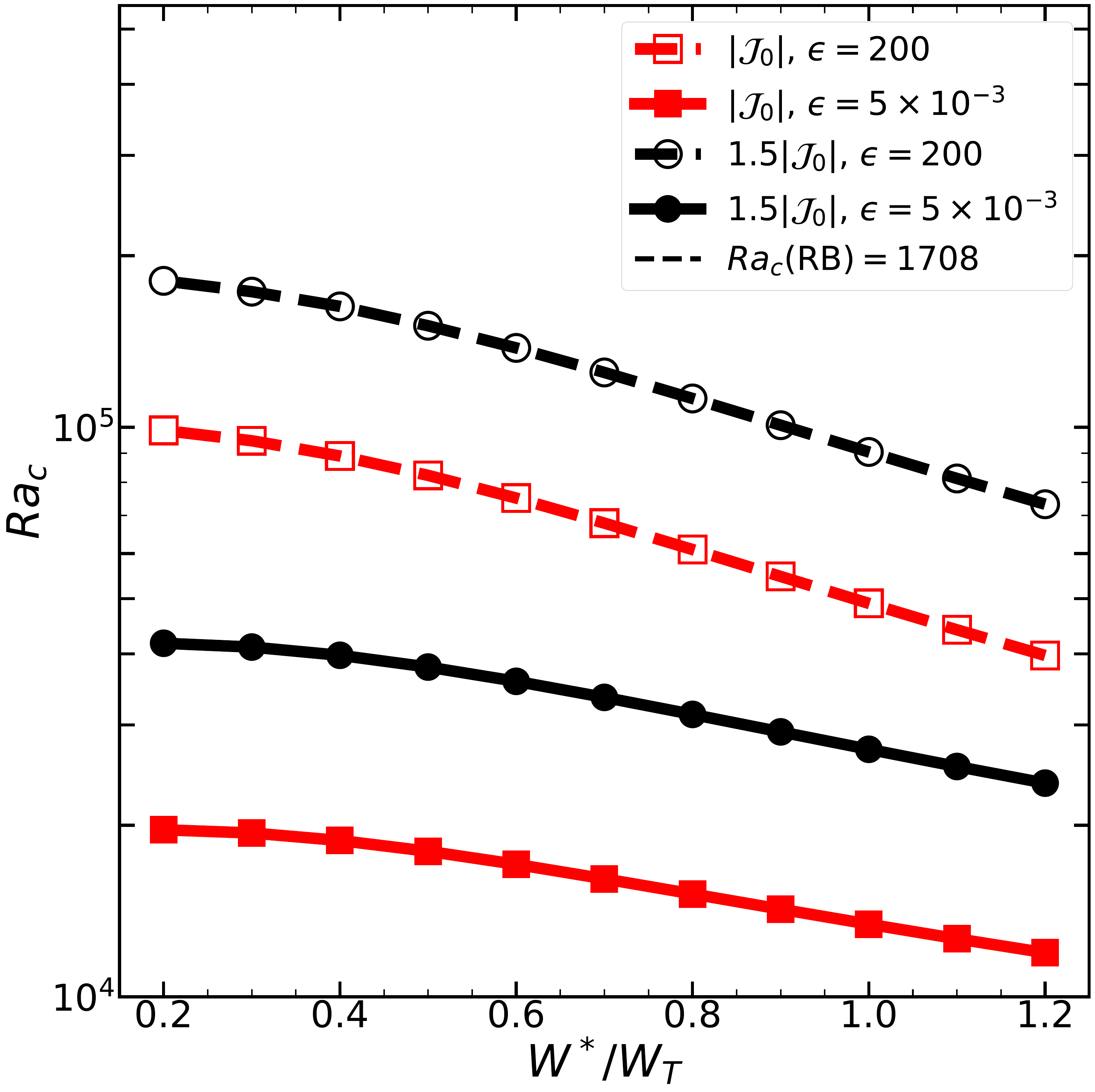}
        \caption{}
        \label{fig:3a_rc}
    \end{subfigure}%
    \hfill
    \begin{subfigure}[t]{0.48\textwidth}
        \centering
        \includegraphics[width=\linewidth]{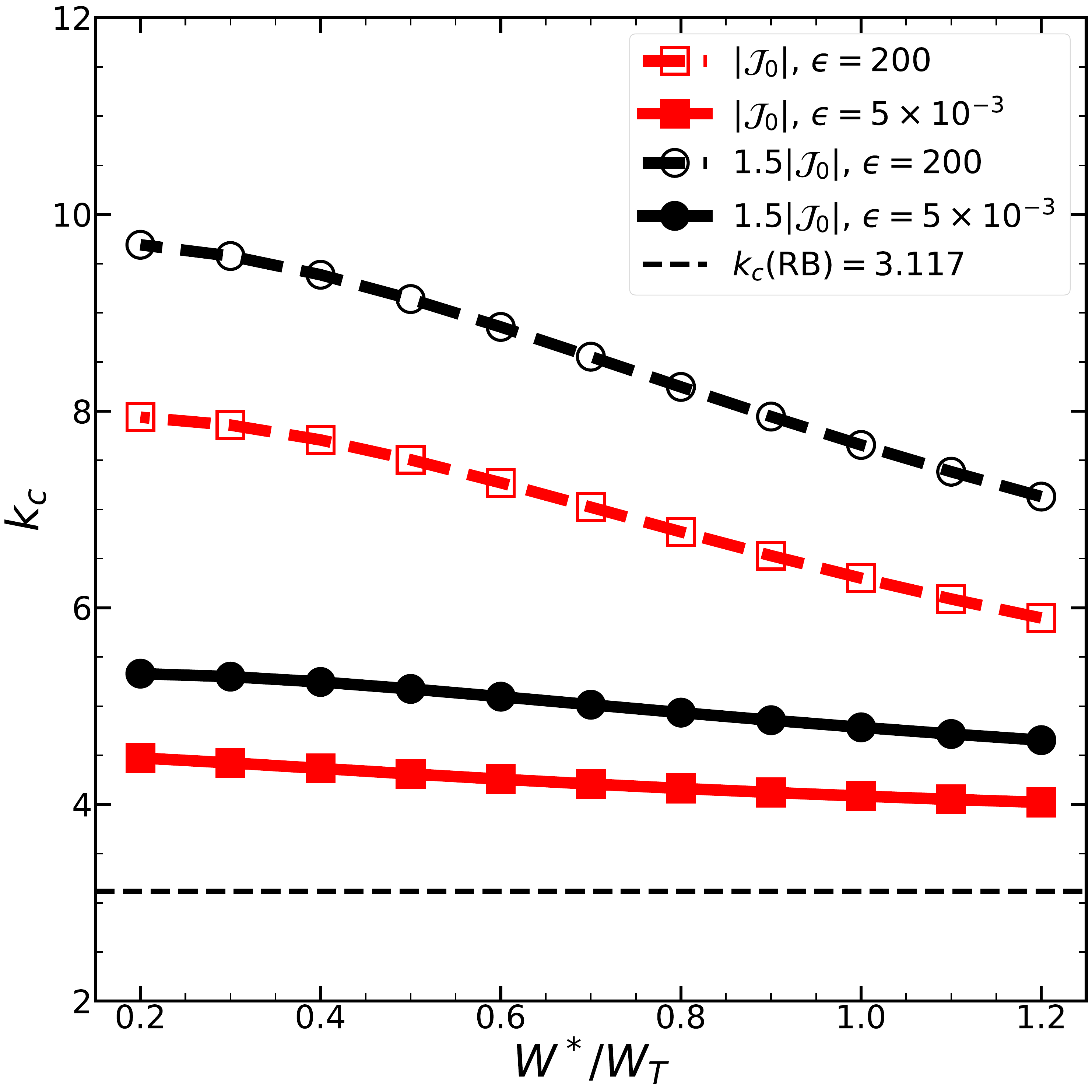}
        \caption{}
        \label{fig:3a_kc}
    \end{subfigure}

    \caption{(a) Critical Rayleigh number and (b) critical wavenumber as functions of inlet velocity, showing the onset of instability for heavy particles with  $\beta = 0.5$ under different thermal coupling strengths and particulate flux conditions.}
    \label{fig:3a}
\end{figure}

\begin{figure}[!htb]
    \centering
    \begin{subfigure}[t]{0.48\textwidth}
        \centering
        \includegraphics[width=\linewidth]{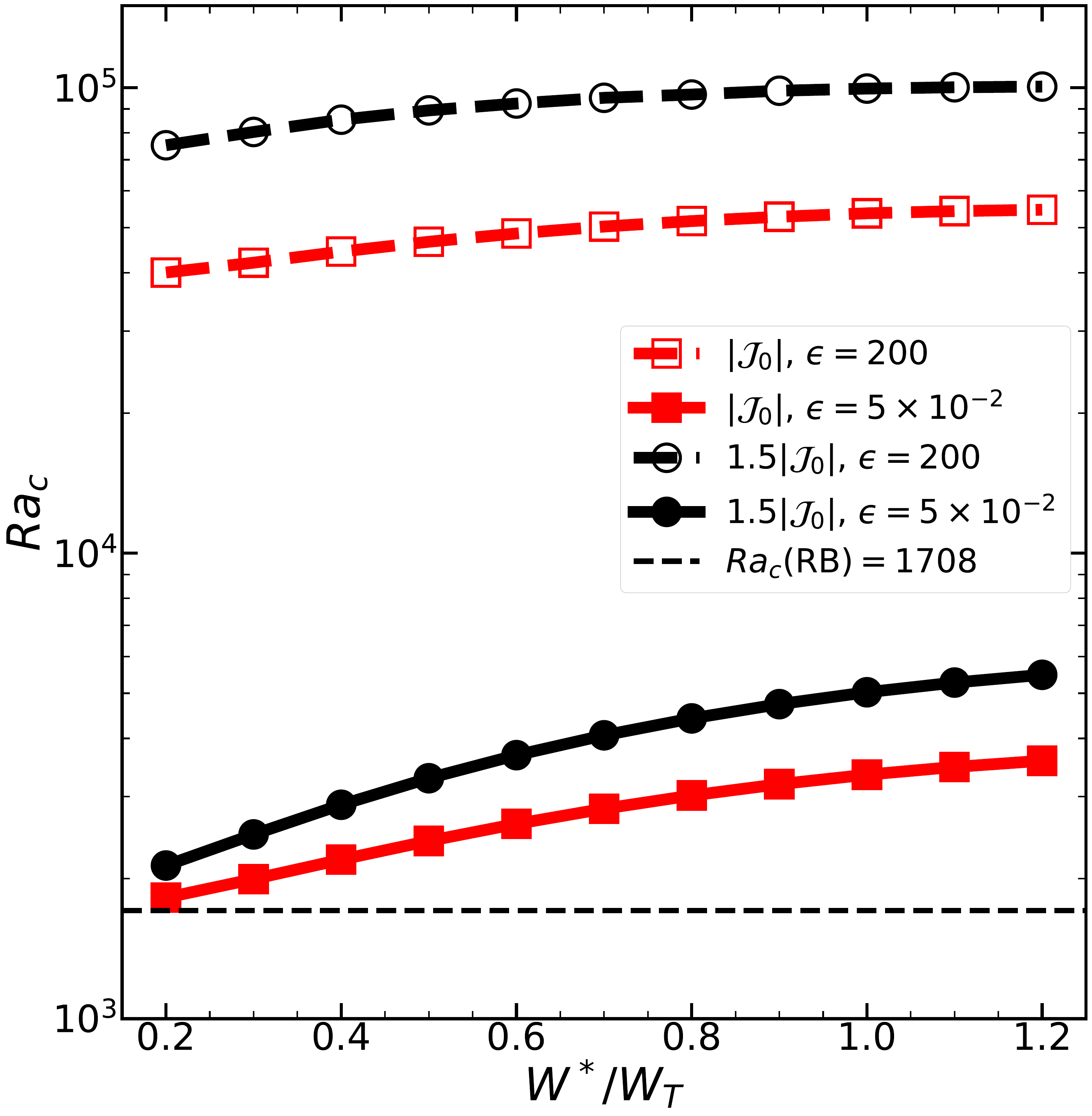}
        \caption{}
        \label{fig:3b_rc}
    \end{subfigure}%
    \hfill
    \begin{subfigure}[t]{0.495\textwidth}
        \centering
        \includegraphics[width=\linewidth]{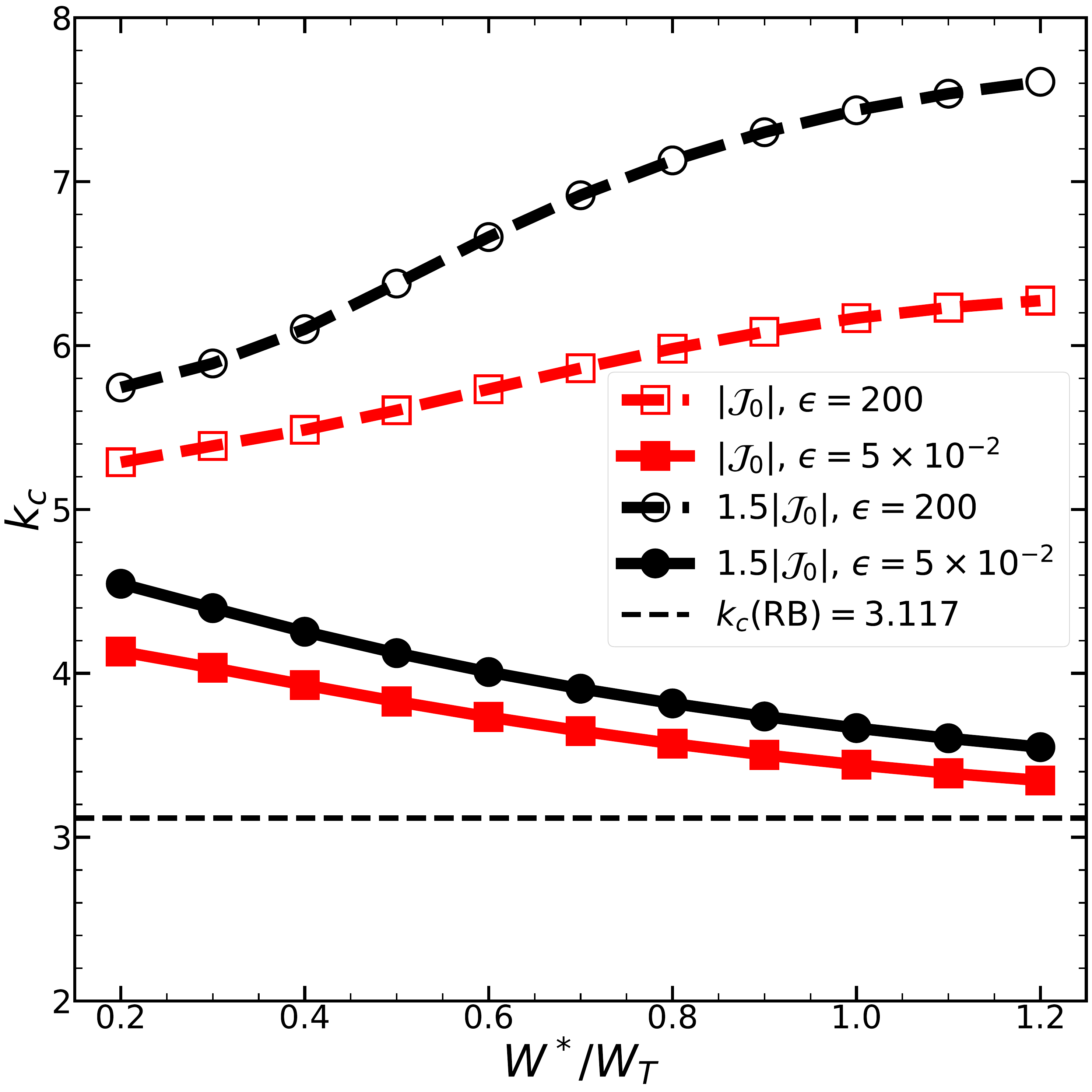}
        \caption{}
        \label{fig:3b_kc}
    \end{subfigure}

    \caption{(a) Critical Rayleigh number ($Ra_c$) and (b) critical wavenumber ($k_c$) as functions of inlet velocity, showing the onset of instability for $\beta = 2.5$ (light particles) under different thermal coupling strengths and particulate flux conditions.}
    \label{fig:3b}
\end{figure}

Both trends discussed above, namely the increase in stability and dominant disturbance size observed when increasing particle thermal inertia, persist when the particle injection velocity $\mathbf{W}^{\ast}$ and volumetric particle flux $\mathcal{J}$ are varied, as respectively illustrated in Figures~\ref{fig:3a} and~\ref{fig:3b}.

For the heavy particles shown in Figure~\ref{fig:3a}, variations in $\mathcal{J}$ and $\mathbf{W}^{\ast}$ do not produce qualitative differences between the weak and strong thermal coupling regimes, corresponding to $\epsilon = 5 \times 10^{-3}$ and $\epsilon = 200$, respectively. However, when the inlet particle flux increases, the influence of $\epsilon$ on the stabilization magnitude becomes strongly amplified: a 50\% increase in $\mathcal{J}$ nearly doubles the value of $Ra_c$. This parameter has a qualitatively similar effect on the dominant wavenumber size $k_c$, although it is not as quantitatively strong.

For the light particles shown in Figure~\ref{fig:3b}, similar trends are observed with respect to $\mathcal{J}$. In contrast, increasing the inlet particle velocity promotes rather than weakens stability. Furthermore, it decreases rather than increases dominant disturbance size for high enough $\epsilon$. Overall, the results show that higher particle fluxes and larger specific heat capacity ratios enhance system stability for both heavy and light particles, although the relative importance of $\mathbf{W}^{\ast}$ differs between the two regimes.

\subsection{Fluid and particle patterns at the onset of instability}

The linear stability analysis enables the visualization of fluid and particle patterns at the onset of natural convection. These fields are representative of the physical system as long as disturbance amplitudes remain small, such that the linear approximation holds, i.e. before nonlinear saturation mechanisms become significant. Despite this limitation, the analysis provides valuable insight into the linear interaction between fluid and particle patterns. In addition, it allows examination of whether a nonuniform spatial distribution of particles can develop at the onset of convection.

\begin{figure}[!htb]
     \centering
        \begin{subfigure}[c]{0.48\textwidth}
         \centering
         \includegraphics[width=\textwidth]{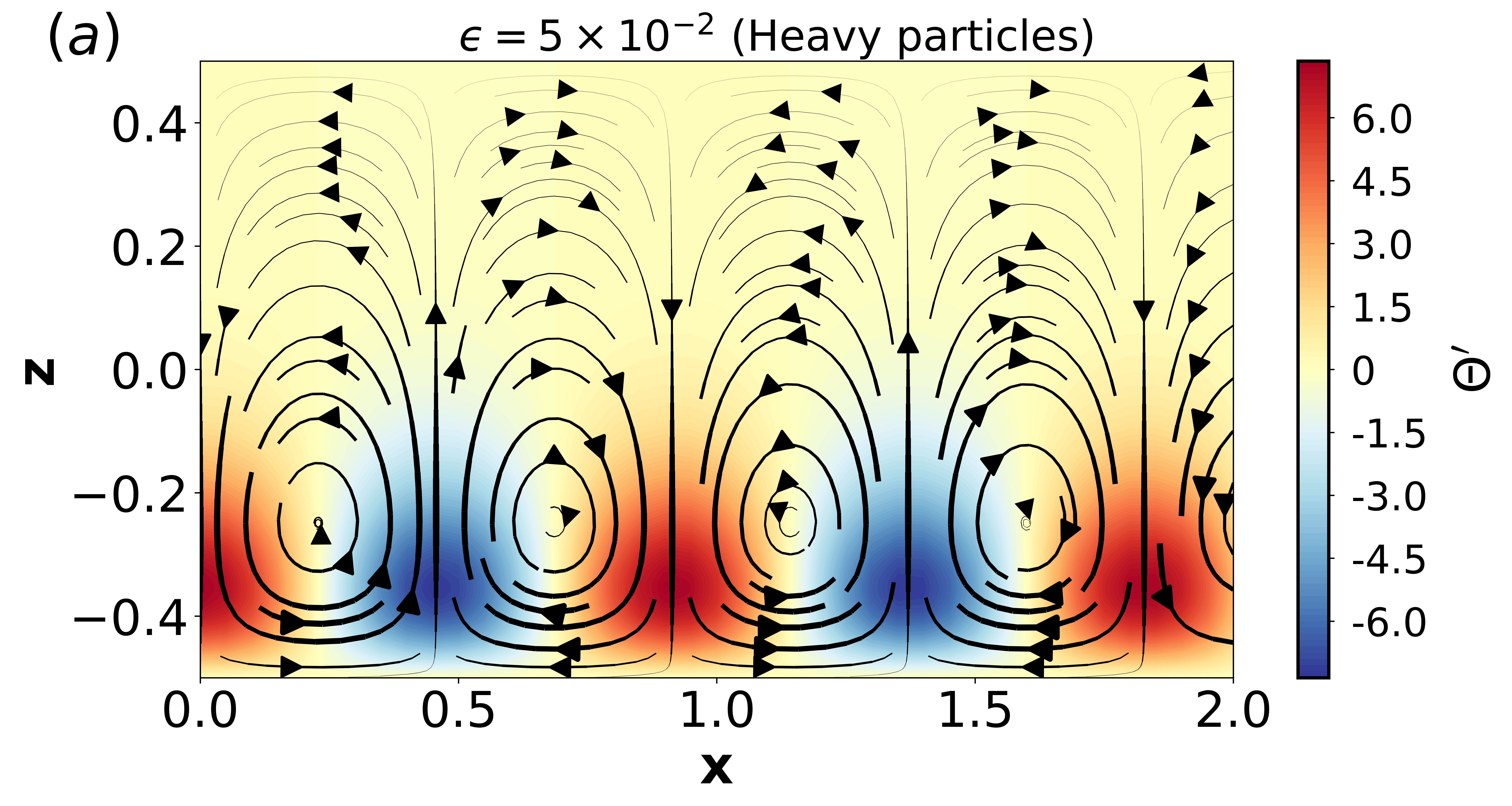}
         \end{subfigure}
              \hfill
              \begin{subfigure}[c]{0.48\textwidth}
         \centering
         \includegraphics[width=\textwidth]{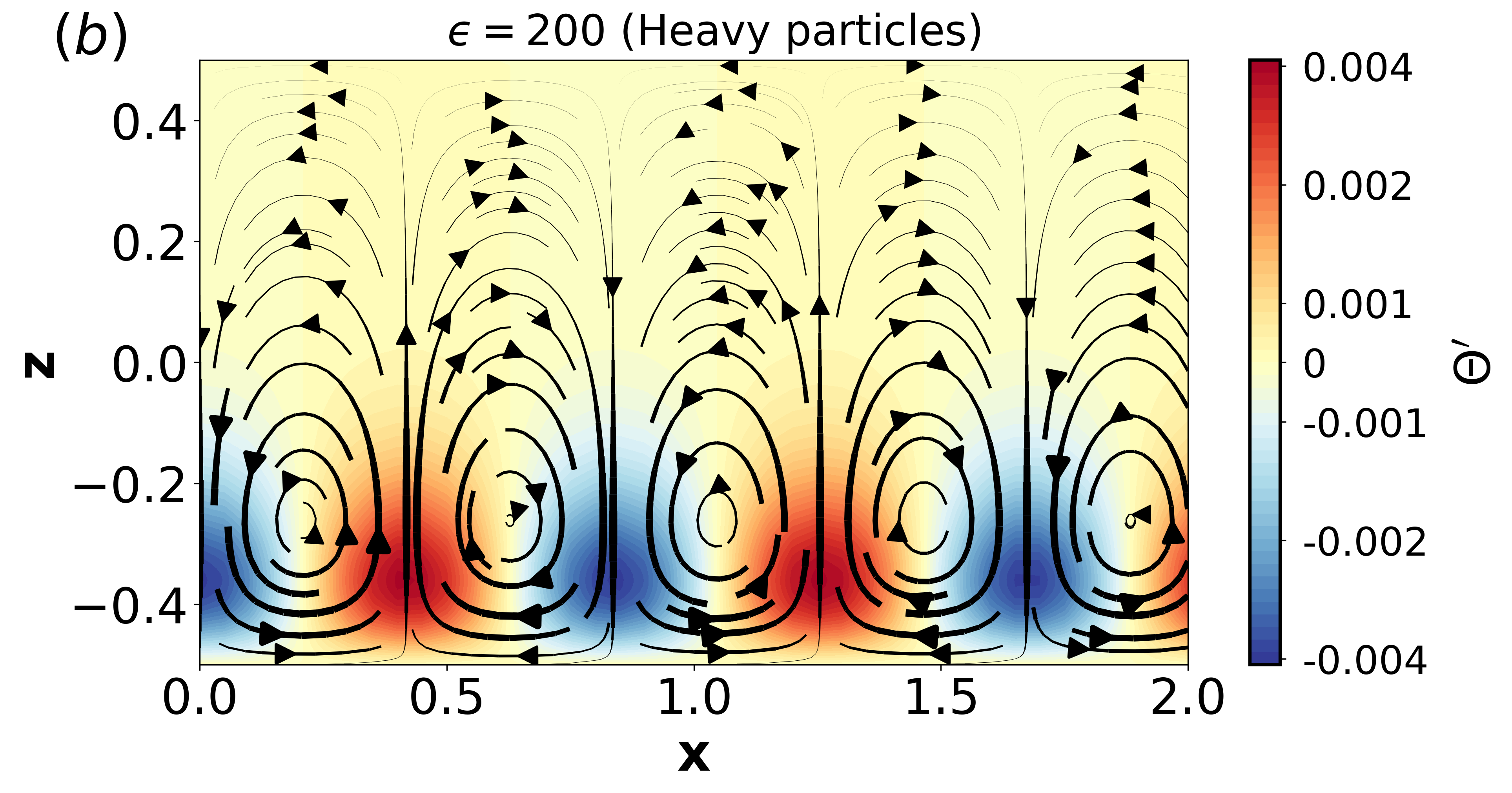}
         \end{subfigure}
              \begin{subfigure}[c]{0.48\textwidth}
         \centering
         \includegraphics[width=\textwidth]{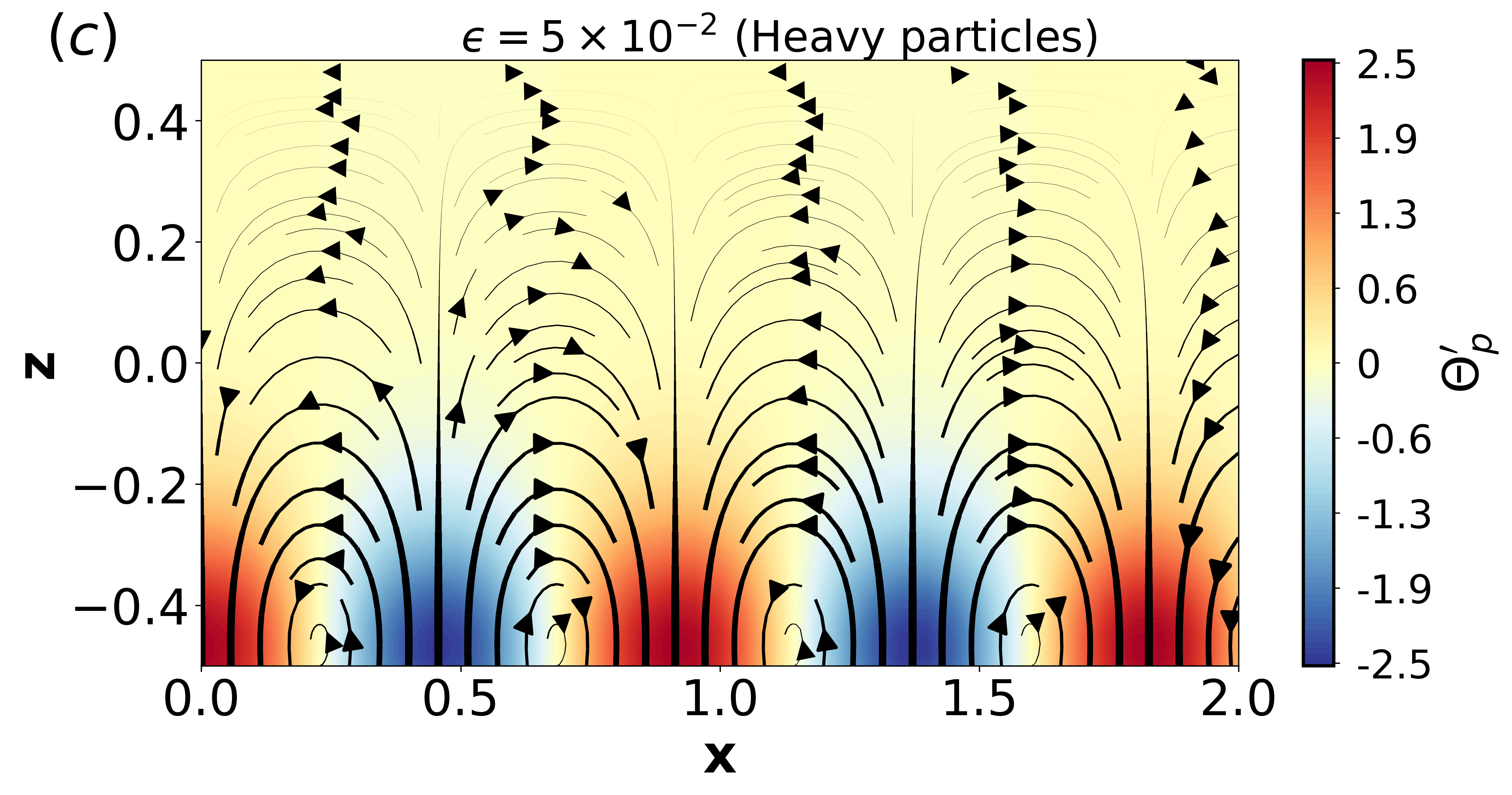}
         \end{subfigure}
     \hfill     
     \begin{subfigure}[c]{0.48\textwidth}
         \centering
         \includegraphics[width=\textwidth]{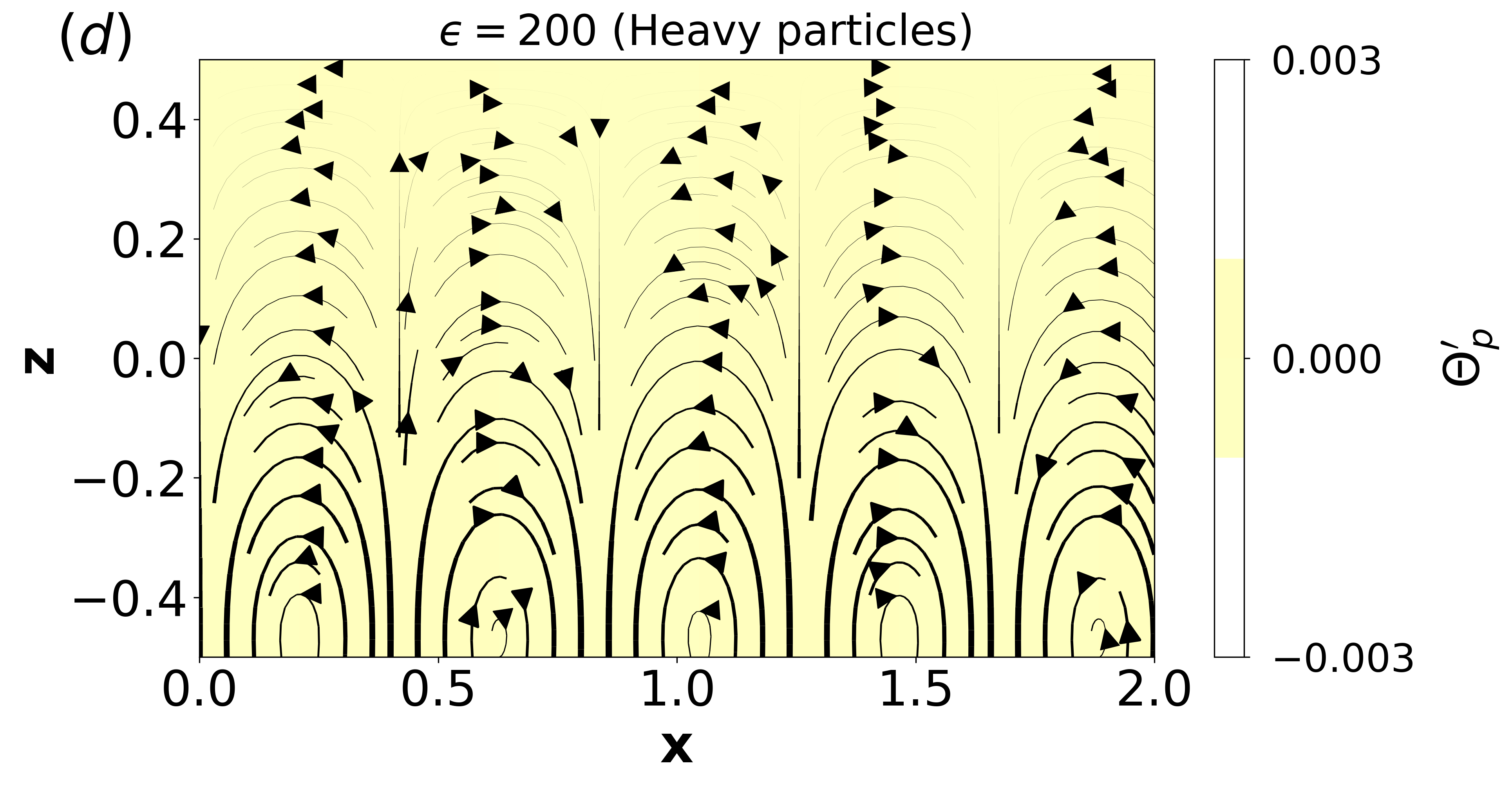}
         \end{subfigure}
              \hfill
     \begin{subfigure}[c]{0.48\textwidth}
         \centering
         \includegraphics[width=\textwidth]{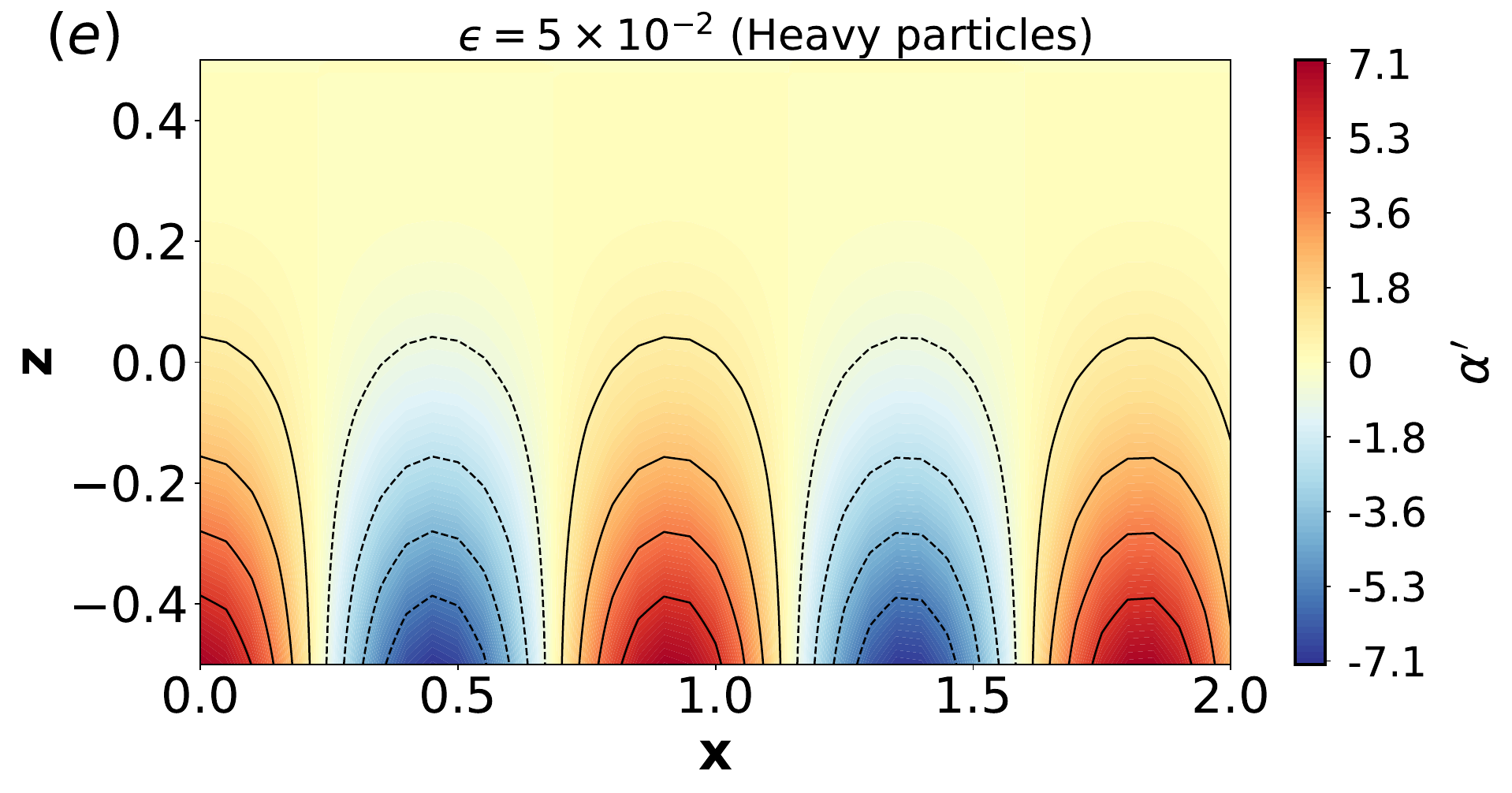}
         \end{subfigure}
     \hfill
     \begin{subfigure}[c]{0.47\textwidth}
         \centering
         \includegraphics[width=\textwidth]{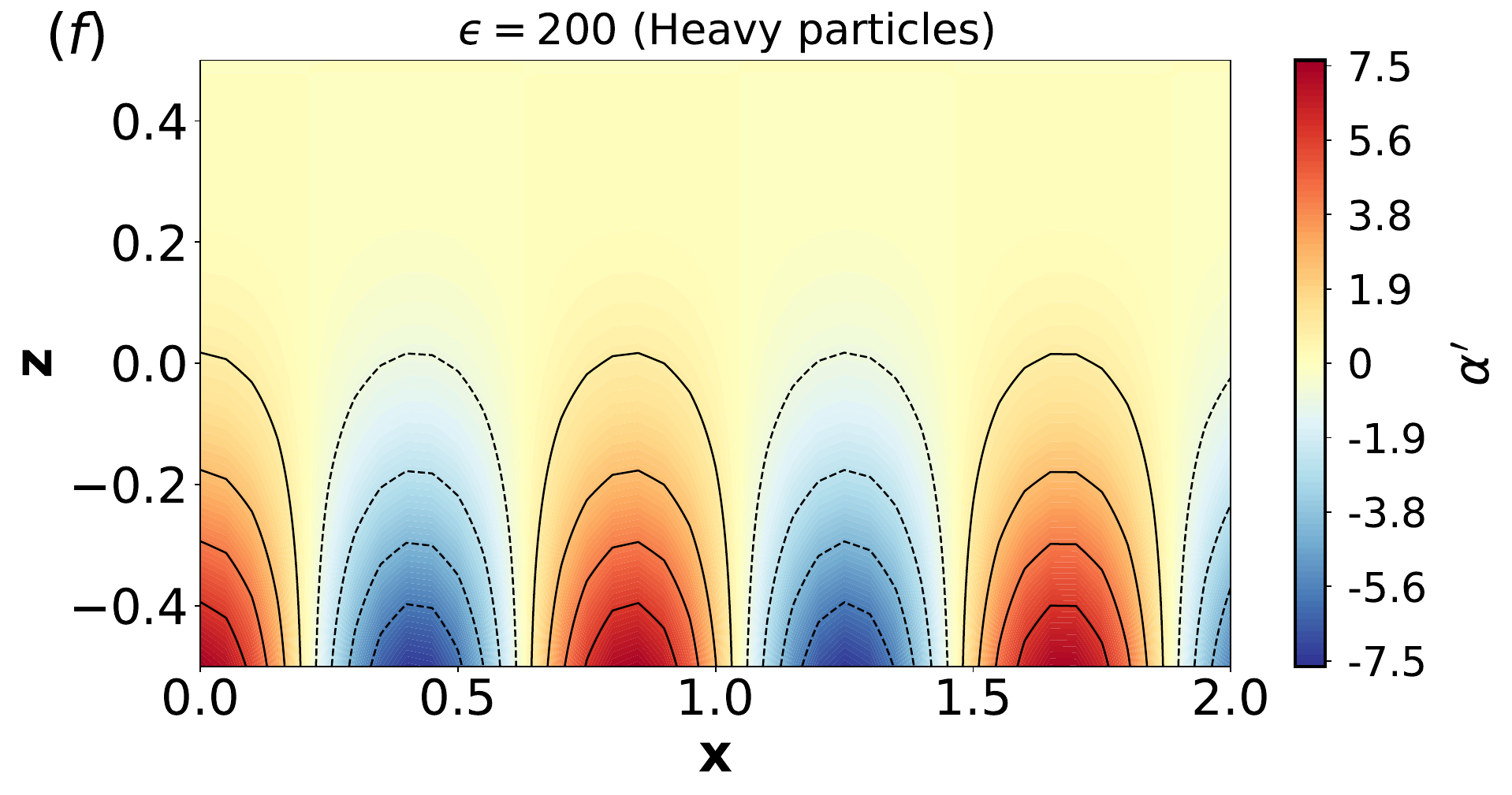}
         \end{subfigure}
              \hfill
     \caption{
Field visualizations for $\beta = 0.5$ (heavy particles), $\mathcal{J}=\mathcal{J}_0$ and $W^*=0.5W_T$: (a-b) Streamlines of the fluid velocity, where the line thickness reflects the local velocity magnitude, overlaid on the fluid temperature field $\Theta'$ (in colors). 
(c-d) Streamlines of the particle velocity field overlaid on the particle temperature field $\Theta_p'$.
(e-f) Contour lines and heatmap for concentration of the particle.}
\label{fig:fig4}
\end{figure}

Within the framework of linearized dynamics, only particle injections with velocities differing from their terminal value can generate spatial inhomogeneities, as shown in \cite{Raza_Freitas_Alves_Calzavarini_Hirata_2025}. Figures~\ref{fig:fig4} present visualizations of the velocity and temperature fields of both the fluid and dispersed phases, together with the particle concentration, at the onset of convection for heavy particles ($\beta = 0.5$). Only disturbance fields are displayed; base-state contributions are omitted.

\begin{figure}[!htb]
     \centering
        \begin{subfigure}[c]{0.48\textwidth}
         \centering
         \includegraphics[width=\textwidth]{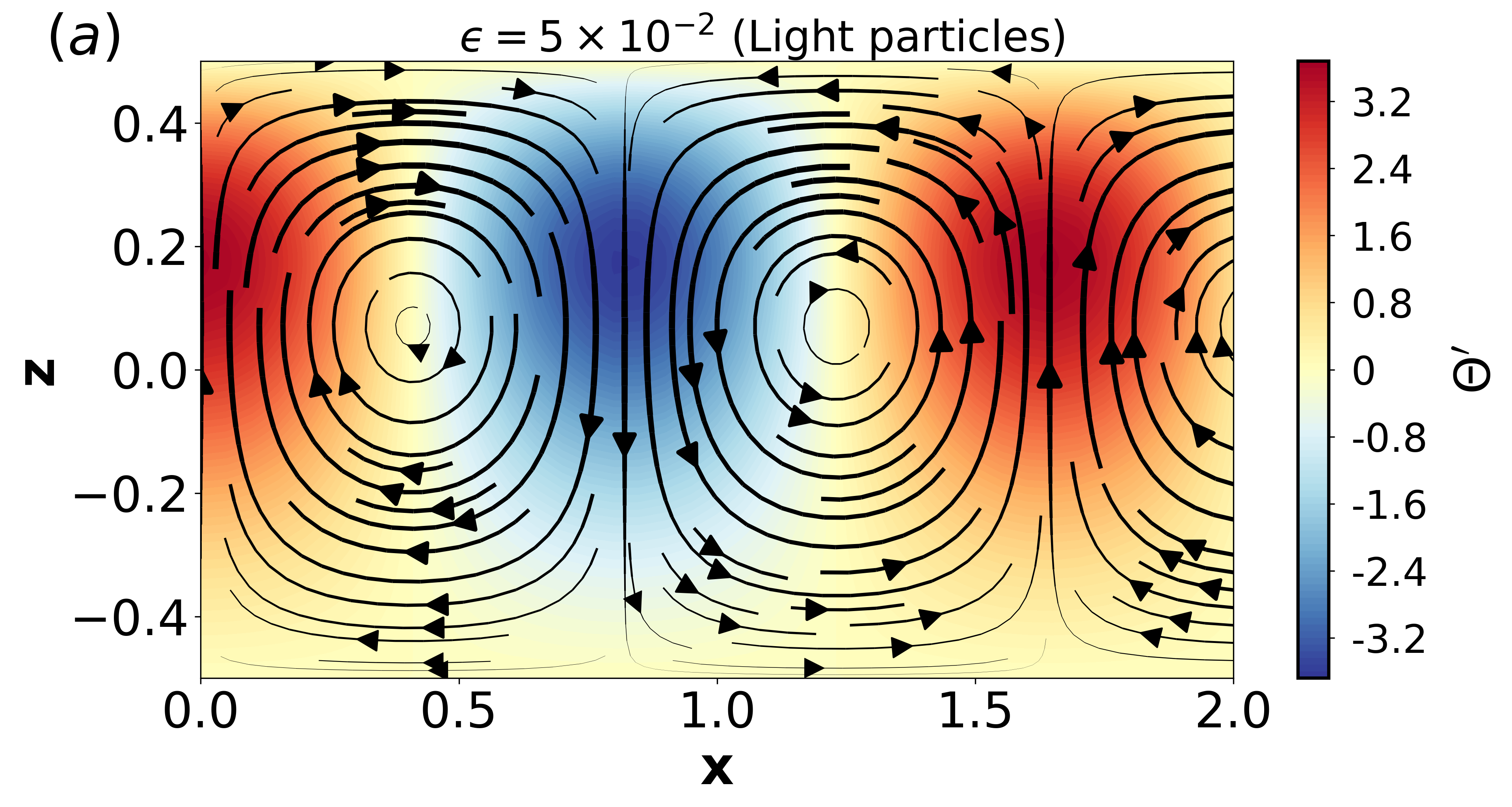}
         \end{subfigure}
              \hfill
     \begin{subfigure}[c]{0.48\textwidth}
         \centering
         \includegraphics[width=\textwidth]{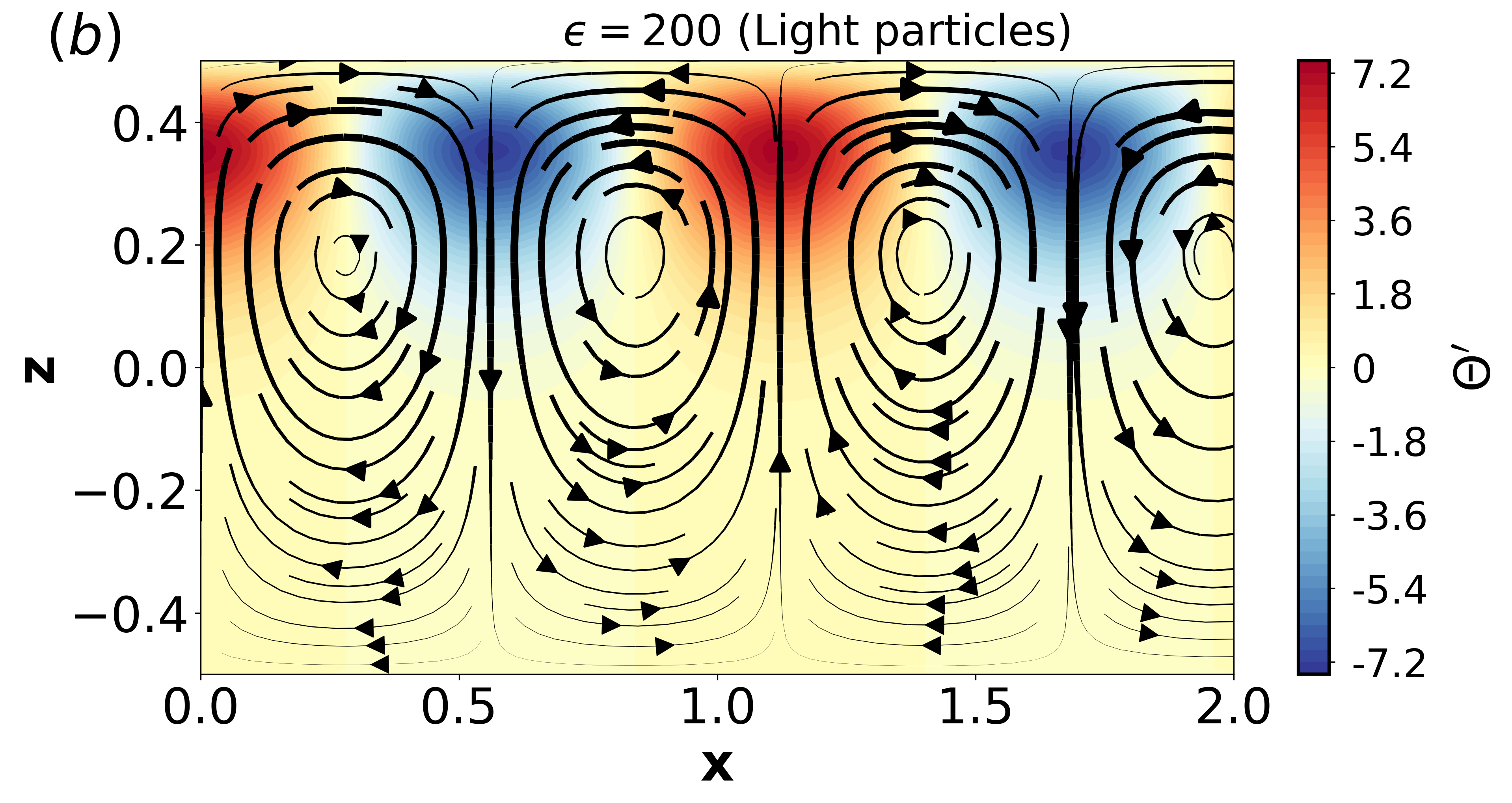}
         \end{subfigure}
              \begin{subfigure}[c]{0.48\textwidth}
         \centering
         \includegraphics[width=\textwidth]{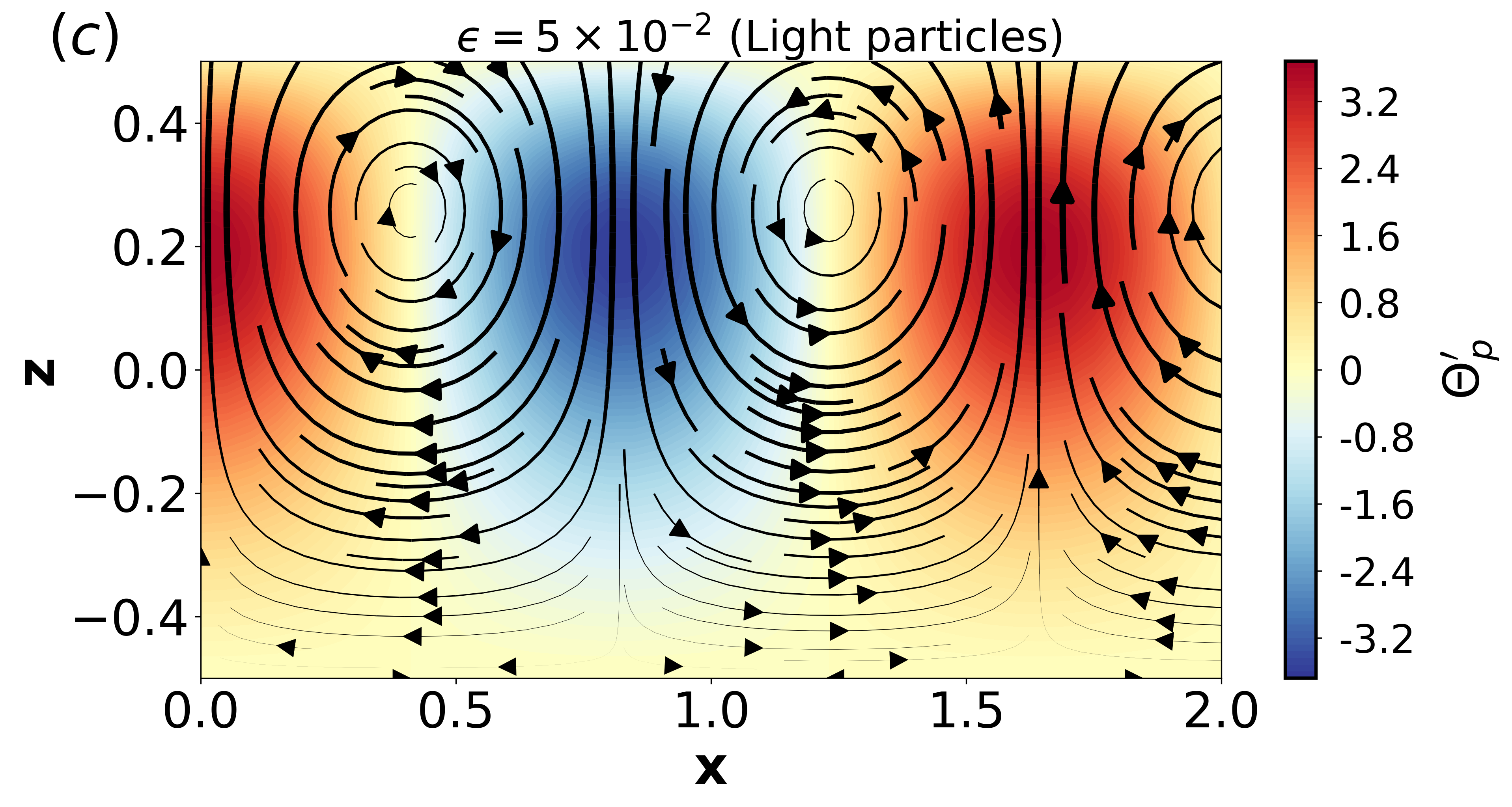}
         \end{subfigure}
     \hfill     
     \begin{subfigure}[c]{0.48\textwidth}
         \centering
         \includegraphics[width=\textwidth]{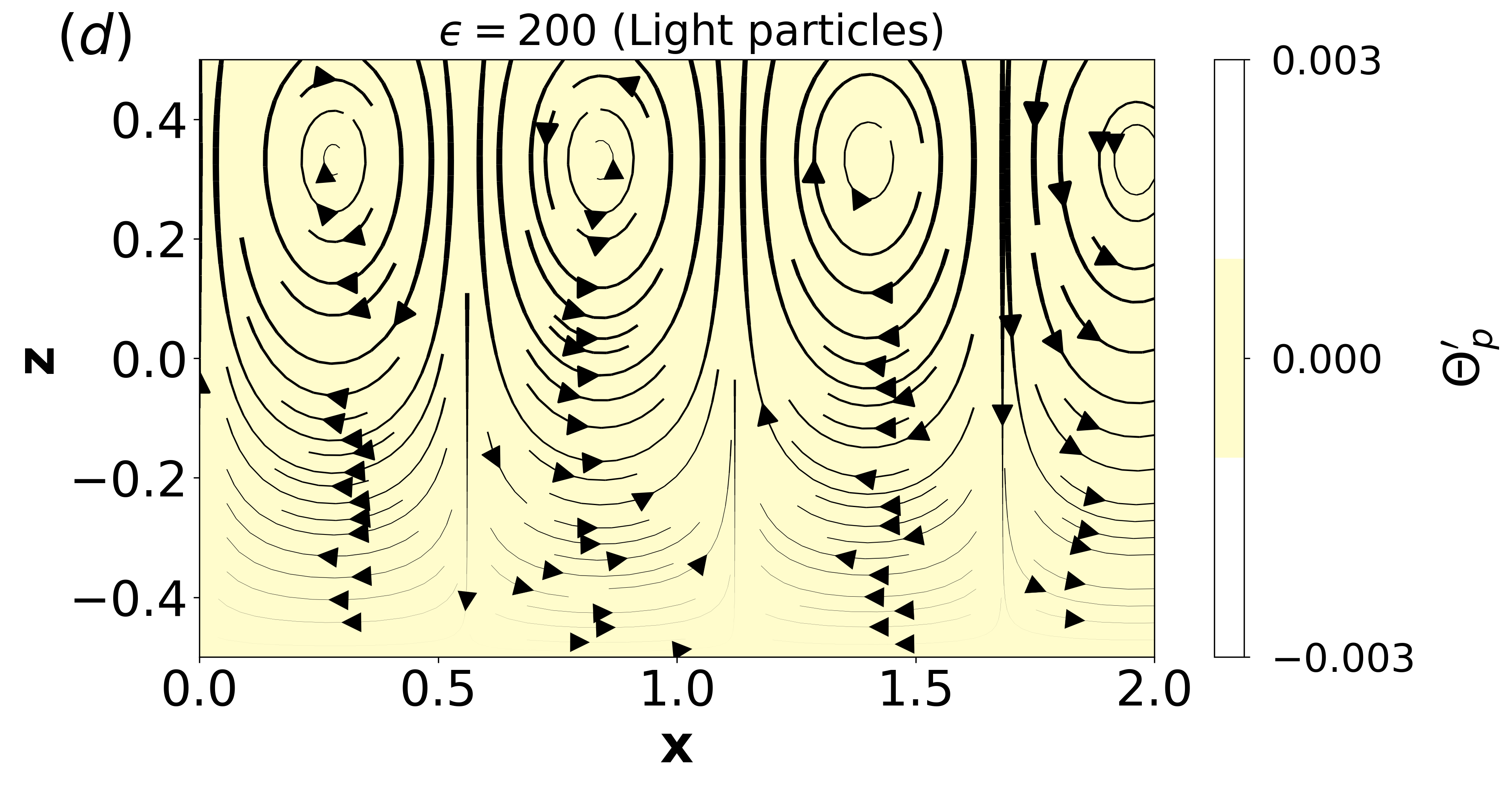}
         \end{subfigure}
              \hfill
     \begin{subfigure}[c]{0.48\textwidth}
         \centering
         \includegraphics[width=\textwidth]{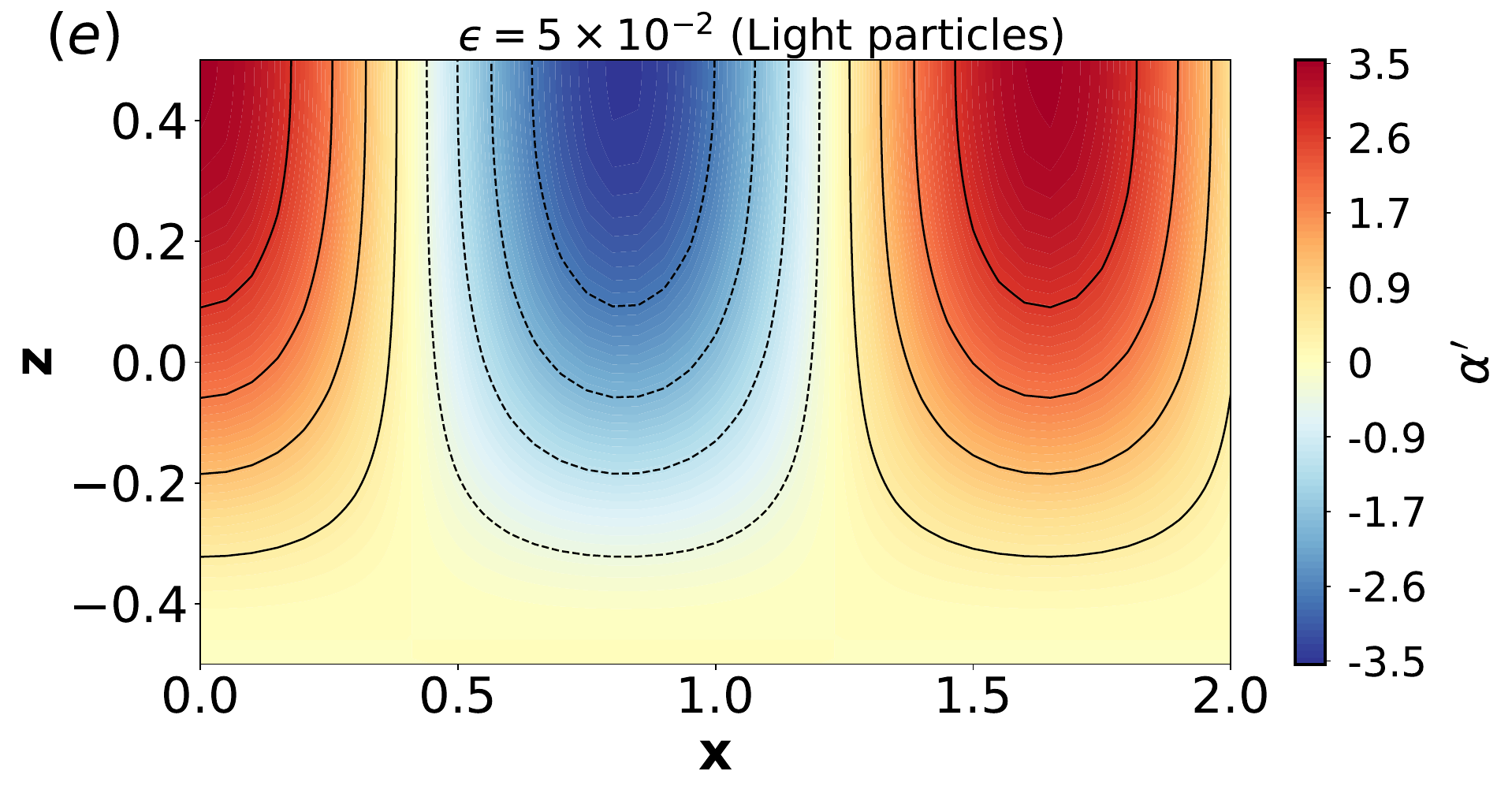}
         \end{subfigure}
     \hfill
     \begin{subfigure}[c]{0.47\textwidth}
         \centering
         \includegraphics[width=\textwidth]{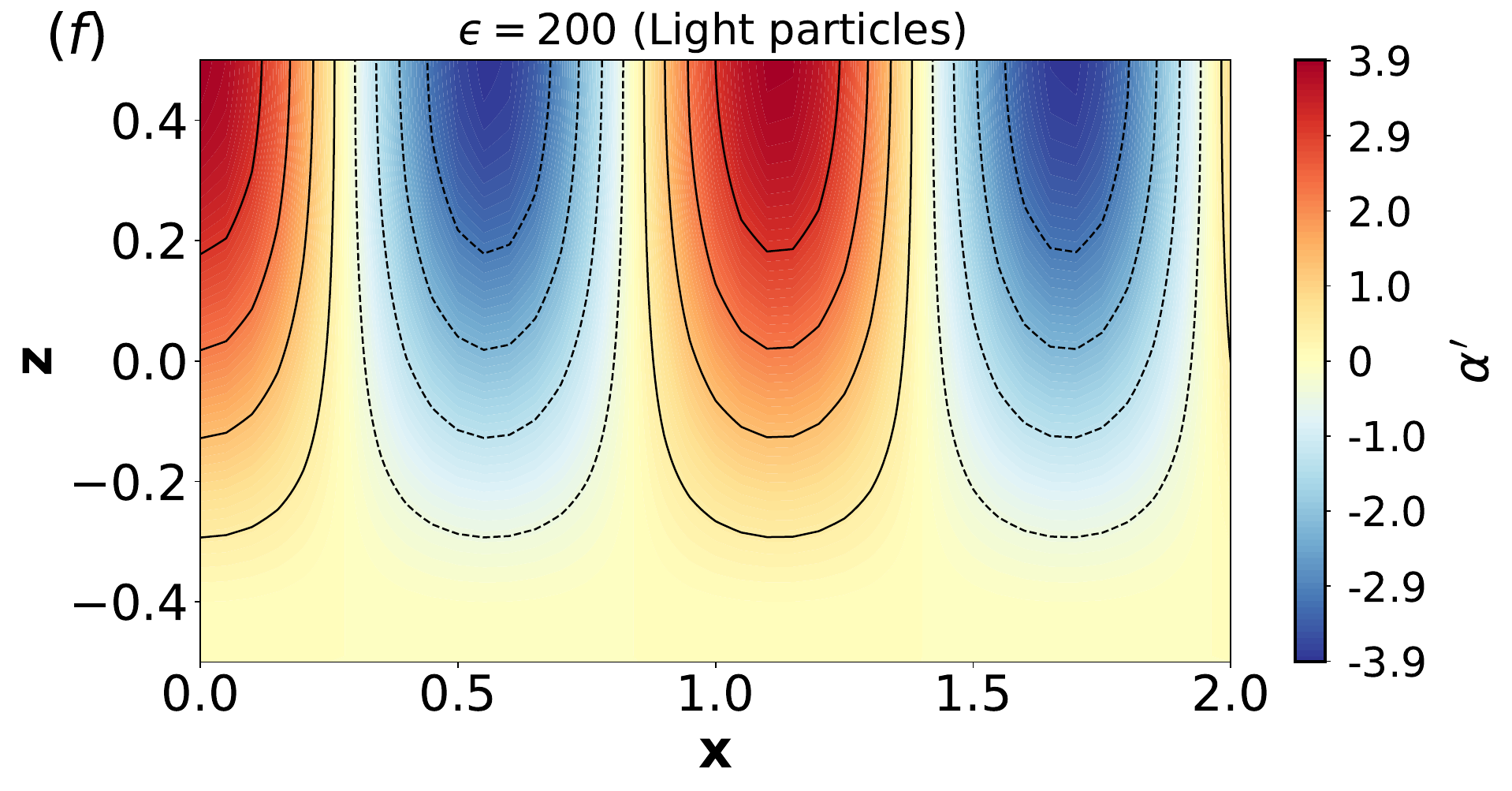}
         \end{subfigure}
              \hfill
       \caption{
Field visualizations for $\beta = 2.5$ (light particles), $\mathcal{J}=\mathcal{J}_0$ and $W^*=0.5W_T$: (a-b) Streamlines of the fluid velocity overlaid on the fluid temperature field $\Theta$  (in colors). 
(c-d) Streamlines of the particle velocity field overlaid on the particle temperature field $\Theta_p'$ (in colors).
(e-f) Contour lines and heatmap for concentration of the particle.}
\label{fig:fig5}
\end{figure}

When particle thermal inertia is high, the particle temperature does not relax to that of the surrounding fluid and therefore remains nearly constant. Regions where both the fluid and particles are colder—that is, exhibit negative temperature disturbances—correspond to zones of high particle concentration. This behavior is expected since initially cold particles are injected from above and dispersed near the {lower wall due to downwelling plumes}. Varying the parameter $\epsilon$ does not alter this trend. Another notable feature is the dependence of the dominant disturbance wavelength on particle inertia: larger inertia leads to smaller convection patterns. Similar behavior is observed for light particles ($\beta = 2.5$), shown in Figure~\ref{fig:fig5}, the only difference being that light particles accumulate near the upper wall in upwelling plumes~\citep{Raza_Freitas_Alves_Calzavarini_Hirata_2025}.

\section{\label{sec:conclusion}Conclusion}

A linear stability analysis of particulate Rayleigh–Bénard convection was conducted to quantify the influence of particle thermal inertia, density ratio, and injection conditions on the onset of convection. The study combined the coupled momentum and energy equations for the fluid and dispersed phases to determine how each parameter modifies the critical Rayleigh number and wavenumber. Particle thermal inertia, represented by the specific heat capacity ratio $\epsilon$, has a stabilizing effect on the system. Increasing $\epsilon$ homogenizes the temperature field, weakens buoyancy-driven motion, and raises both the critical Rayleigh number and the dominant wavenumber, indicating smaller convection cells. This stabilizing trend is independent of the particle injection temperature and roughly saturates beyond $\epsilon = \mathcal{O}(1)$. The density ratio $\beta$ exerts an asymmetric influence: both heavy ($\beta < 1$) and light ($\beta > 1$) particles stabilize the system relative to the single-phase case, but the effect is stronger for heavy particles. For $\beta \to 1$, the model reaches a singular limit where maintaining a constant particle flux would violate the dilute suspension assumption. For large $\beta$, the system approaches the “bubble” limit ($\beta = 3$), where thermal coupling vanishes. In such case the system may become even more unstable that the classical single phase RB system. Variations in the inlet particle velocity $\mathbf{W}^{\ast}$ and volumetric flux $\mathcal{J}$ modulate the stabilizing effect of $\epsilon$. Higher fluxes enhance the influence of particle thermal inertia for both heavy and light particles. However, increasing $\mathbf{W}^{\ast}$ promotes stability in the light-particle regime but not in the heavy-particle one.

These findings clarify the distinct roles of particle thermal and mechanical couplings in the stability of particulate Rayleigh–Bénard systems. They provide a quantitative foundation for nonlinear analysis either theoretical or numerical and experimental studies, which are currently being pursued. \\

\textbf{Declaration of interests.} The authors report no conflict of interest.\\

\textbf{Funding.} This work was partially supported by the CAPES-COFECUB programme (project number: Ph1097/26), funded by the French Ministry for Europe and Foreign Affairs, the French Ministry for Higher Education and CAPES.

\clearpage

\end{document}